\documentclass[twocolumn]{aastex63}
\usepackage{amsmath,amstext}
\usepackage[T1]{fontenc}
\usepackage{apjfonts} 
\usepackage[figure,figure*]{hypcap}
\usepackage{scalerel,amssymb}

\newcommand{\kmin}{k_{\rm min}}
\newcommand{\kmax}{k_{\rm max}}
\newcommand{\kc}{k_{\rm c}}
\newcommand{\lrot}{\lambda_{\rm rot}}
\newcommand{\lrotc}{\lambda_{\rm rot, c}}
\newcommand{\orot}{\omega_{\rm rot}}
\newcommand{\lj}{\lambda_{\rm J}}

\newcommand{\lequi}{\lambda _{\rm eq}}
\newcommand{\lk}{\lambda_c}
\newcommand{\erf}{{\rm erf}}
\newcommand{\msun}{\rm M_{\odot}}
\newcommand{\btheta}{\boldsymbol{\theta}}
\newcommand{\bomega}{\boldsymbol{\omega}}
\newcommand{\bvel}{\pmb{v}}

\newcommand{\mV}{\mathcal{V}}

\newcommand{\ono}{\omega_{\rm nc}}

\newcommand{\gno}{\gamma_{\rm nc}}
\newcommand{\grot}{\gamma_{\rm rot}}
\newcommand{\vno}{v_{\rm nc}}
\newcommand{\vrot}{v_{\rm rot}}

\shorttitle{Characterizing ISM Circulation in Disk Galaxies}
\shortauthors{Utreras et al.}

\begin{document}

\title{When Gas Dynamics Decouples from Galactic Rotation: Characterizing ISM Circulation in Disk Galaxies}
\correspondingauthor{Jos\'e Utreras}
\email{jutreras@ug.uchile.cl}

\author{Jos\'e Utreras}
\affiliation{Departamento de Astronom\'ia, Universidad de Chile, Casilla 36-D, Santiago, Chile}

\author{Guillermo A. Blanc}
\affiliation{Observatories of the Carnegie Institution for Science, 813 Santa Barbara Street, Pasadena, CA 91101, USA}
\affiliation{Departamento de Astronom\'ia, Universidad de Chile, Casilla 36-D, Santiago, Chile}

\author{Andr\'es Escala}
\affiliation{Departamento de Astronom\'ia, Universidad de Chile, Casilla 36-D, Santiago, Chile}

\author{Sharon Meidt}
\affiliation{Sterrenkundig Observatorium, Universiteit Gent, Krijgslaan 281 S9, B-9000 Gent, Belgium}

\author{Eric Emsellem}
\affiliation{European Southern Observatory, Karl-Schwarzschild-Str. 2, D-85748 Garching, Germany}
\affiliation{Universit\'e Lyon 1, ENS de Lyon, CNRS, Centre de Recherche Astrophysique de Lyon UMR5574, F-69230 Saint-Genis-Laval, France}

\author{Frank Bigiel}
\affiliation{Argelander-Institut f\"ur Astronomie, Universit\"at Bonn, Auf dem H\"ugel 71, D-53121 Bonn, Germany}

\author{Simon ~C.~O.~Glover}

\affiliation{Zentrum f\"{u}r Astronomie, Universit\"{a}t Heidelberg, Institut f\"{u}r Theoretische Astrophysik,  Albert-Ueberle-Str. 2, D-69120 Heidelberg, Germany}

\author{Jonathan Henshaw}
\affiliation{Max-Planck-Institut f\"ur Astronomie, K\"onigstuhl 17, D-69117 Heidelberg, Germany}

\author{Alex Hygate}
\affiliation{Max-Planck-Institut f\"ur Astronomie, K\"onigstuhl 17, D-69117 Heidelberg, Germany}
\affiliation{Astronomisches Rechen-Institut, Zentrum f{\"u}r Astronomie der Universit{\"a}t Heidelberg, M{\"o}nchhofstra{\ss}e 12-14, D-69120 Heidelberg, Germany}

\author{J.~M.~Diederik Kruijssen}
\affiliation{Astronomisches Rechen-Institut, Zentrum f{\"u}r Astronomie der Universit{\"a}t Heidelberg, M{\"o}nchhofstra{\ss}e 12-14, D-69120 Heidelberg, Germany}

\author{Erik Rosolowsky}
\affiliation{Department of Physics, 4-181 CCIS, University of Alberta, Edmonton, AB T6G 2E1, Canada}
\author{Eva Schinnerer} 
\affiliation{Max-Planck-Institut f\"ur Astronomie, K\"onigstuhl 17, D-69117 Heidelberg, Germany}
\author{Andreas Schruba}
\affiliation{Max-Planck Institut f\"ur Extraterrestrische Physik, Giessenbachstra{\ss}e 1, D-85748 Garching, Germany}

\begin{abstract}

In galactic disks, galactic rotation sets the bulk motion of gas, and its energy and momentum can be transferred toward small scales. Additionally, in the interstellar medium, random and noncircular motions arise from stellar feedback, cloud-cloud interactions, and instabilities, among other processes. Our aim is to comprehend to what extent small-scale gas dynamics is decoupled from galactic rotation. We study the relative contributions of galactic rotation and local noncircular motions to the circulation of gas, $\Gamma$, a macroscopic measure of local rotation, defined as the line integral of the velocity field around a closed path. We measure the circulation distribution as a function of spatial scale in a set of simulated disk galaxies and we model the velocity field as the sum of galactic rotation and a Gaussian random field. The random field is parameterized by a broken power law in Fourier space, with a break at the scale $\lk$. We define the spatial scale $\lequi$ at which galactic rotation and noncircular motions contribute equally to $\Gamma$. For our simulated galaxies, the gas dynamics at the scale of molecular clouds is usually dominated by noncircular motions, but in the center of galactic disks galactic rotation is still relevant. Our model shows that the transfer of rotation from large scales breaks at the scale $\lk$ and this transition is necessary to reproduce the circulation distribution. We find that $\lequi$, and therefore the structure of the gas velocity field, is set by the local conditions of gravitational stability and stellar feedback. 

\end{abstract}


\section{Introduction}

The structure of the gas velocity field is crucial for understanding how galaxies and molecular clouds evolve. The dynamical state of gas is one of the key elements in star formation theories, e.g. invoking turbulence at the scale of molecular clouds \citep{Padoan_12,Semenov_15} or galactic rotation as a particular parameter controlling star formation at galactic scales \citep{Elmegreen_97,Silk_97,Kennicutt_98,Tan_00,Krumholz_12,Utreras_2016, Jeffreson_18,Meidt_2018}. The common picture of star formation involves self-gravity and sources of energy acting against self-gravity. Galactic rotation is one of those energy sources, acting at the largest spatial scales, where ordered motions make up the bulk of the kinetic energy \citep{Utreras_2016,Colling_18,Meidt_2020}. However, while its importance is evident on large scales, it is not clear down to which spatial scales galactic rotation remains dynamically relevant. At the scales of molecular clouds or stellar cores, gas can be dynamically less coupled with galactic rotation, and local noncircular motions start to dominate.

These nonordered or noncircular motions are originated by gravitational instabilities, hydrodynamical instabilities \citep{Matsumoto_10,Renaud_13,Sormani_17}, cloud-cloud interactions, torques from nonaxisymmetric potentials, gas accretion, and stellar feedback \citep{Goldbaum_15,Krumholz_2018}. These energy sources inject turbulence and induce noncircular motions that cascade toward small and large scales \citep{Kraichnan_67,Bournaud_10}. The scale at which gas motion goes from being dominated by ordered rotation to these noncircular dynamical regimes depends on the importance of these other processes relative to ordered rotation. While large-scale dynamics are set by the galactic angular velocity $\Omega(R)=V(R)/R$, small-scale noncircular motions are more difficult to model. 

At galactic scales many studies have focused on the role of galactic rotation in the stabillity of gaseous rotating disks described by the Toomre parameter $Q$ \citep{Toomre_63}. If $Q<1$, the disk is gravitationally unstable to radial perturbations. The classical form of this parameter involves the stability of a razor-thin disk of gas with $Q=\frac{\kappa c_s}{\pi G \Sigma_{\rm gas}}$ where $c_s$ is the gas sound speed,$\kappa$ is the epicyclic frequency given by $\kappa^2=4\Omega^2\left(1+\frac{1}{2}\frac{\partial\ln \Omega}{\partial\ln R} \right)$, $\Sigma_{\rm gas} $ is the gas surface density, and $G$ is the gravitational constant. This ideal case illustrates how galactic rotation delivers support against collapse, in particular to perturbations of size $\lambda>\lambda_{\rm rot}\equiv 4\pi^2 G\Sigma_{\rm gas} /\kappa^2$, setting a maximum size of collapsing fragments \citep{Escala_08}.  Following this body of work, it is natural to expect that rotation plays a significant role in the dynamics of sufficiently large molecular clouds and ultimately in the process of star formation on galactic scales. In particular, the works of \cite{Padoan_12} and \cite{Utreras_2016} show a difference in the efficiency of star formation at these two different scales. \cite{Padoan_12} found that in a turbulent cloud the efficiency is proportional to $\exp({-t_{\rm ff}/t_{\rm cr}})$, while at galactic scales \cite{Utreras_2016} found an efficiency proportional to $\exp({-t_{\rm ff}/t_{\rm orb}})$, where $t_{\rm ff}$, is the initial freefall time, $t_{\rm cr}$, is the cloud crossing time, and $t_{\rm orb}=2\pi/\Omega$ is the orbital time. These works invoke different processes as being important to control the star formation process for two different spatial regimes. It is expected that dynamics are linked from large to small scales; however, most observational studies and theories have neglected the multiscale nature of gas dynamics.\\

As we move our analysis toward the scale of molecular clouds, noncircular motions start to become relevant. A significant body of observational and theoretical research has been devoted to studying the balance between gravitational potential energy $W$ and kinetic energy $K$, commonly described by the virial parameter $\alpha_{\rm vir} = 2K/W$ \citep{Leroy_2017,Padoan_17,Sun_2018}. CO measurements in the PHANGS-ALMA\footnote{Physics at High Angular-resolution in Nearby GalaxieS with ALMA: www.phangs.org} survey made by \cite{Sun_2018} show that $\alpha_{\rm vir}$ varies weakly from cloud to cloud, with $ \alpha_{\rm vir} \sim 1.5 - 3.0$, expected values for marginally bound clouds or free-falling gas. Simulations of turbulent molecular clouds from \cite{Padoan_12} have shown that the efficiency of star formation per freefall time $\epsilon_{\rm ff}$, is sensitive to the strength of self-gravity, and decreases exponentially with $\alpha_{\rm vir}$ (see also \citealp{Federrath_12,Semenov_15}). This anticorrelation between $\alpha_{\rm vir}$ and $\epsilon_{\rm ff}$ has been observed in M51  \citep{Leroy_2017} and in low-pressure atomic-dominated regions in nearby galaxies \citep{Schruba_19}.\\

A common assumption is that at the scale of molecular clouds, most of the kinetic energy $K$ comes from noncircular turbulent motions. This assumption is supported by measurements of velocity gradients of molecular clouds in nearby galaxies. Studying our Galaxy, \cite{Koda_2006} estimated that the fractions of clouds with prograde or retrograde rotation with respect to the Galaxy's spin are similar. Another studied galaxy in this subject is M33: by measuring velocity gradients, \cite{Rosolowsky_2003} found that, if clouds do rotate, nearly 40\% of molecular clouds are counterrotating with respect to the galaxy. More recently, \cite{Braine_2018} found that in M33 molecular clouds do rotate and that their rotation is low, contributing little to the support of the cloud against gravity. These results are expected for clouds dominated by noncircular motions that have randomly aligned spins. In the field of simulations, \cite{Tasker_09} found similar fractions of prograde and retrograde clouds in a simulated Milky Way like galaxy, even in the absence of stellar feedback. \cite{Tasker_09} argued that as time progresses, cloud-cloud interactions inject turbulence at the scale of these interactions.

However, the relevance of galactic rotation versus noncircular motions might depend on the local environment or position in the galactic disk. To compare galactic rotation and noncircular motions, our interest focuses on in-plane motions.  \cite{Meidt_2018} argue that radial variations in the galactic potential are able to influence the dynamics of molecular clouds. Particularly, \cite{Meidt_2018} argue that the internal dynamics of a cloud depend on the velocity field and the cloud size $R_c$ relative to the epicyclic frequency $\kappa$.  For example, if $R_c > \sigma_v/\kappa$, where $\sigma_v$ is the velocity dispersion of gas, the Coriolis force is still relevant in the dynamics of molecular clouds. In other words, for gas structures larger than $ \sigma_v/\kappa$ galactic rotation is still relevant. Since $\kappa$ and $\sigma_v$ vary with galactocentric radius, we might expect that the dynamics of molecular clouds change across a galaxy. Simulations of the galactic center \citep{Kruijssen_19} show that molecular clouds are dominated by strong shear and tidal deformations. Moreover, shear motions from galactic rotation might set the cloud lifetimes in certain conditions \citep{Jeffreson_18}.

One way to estimate the role of galactic rotation is to measure its impact in the local rotation of gas, i.e. the rotation measured with respect to an inertial reference frame. The local rotation of gas is influenced by the large-scale motion of the galaxy and by noncircular motions acting on multiple scales. Our aim is to create a framework that allows us to obtain the contributions from these two types of motion to the local rotation. Since nonordered motions have multiple sources, we need to adopt a statistical approach. A first-order approximation is to consider noncircular motions as a Gaussian random field (GRF), described by a generating function in Fourier space $\mV(k)$, where $k$ is the wavenumber. If we know $\mV(k)$ we also know the magnitude of noncircular motions as a function of spatial scale. Ultimately, $\mV(k)$ let us know at which scales galactic rotation is still relevant. 

We will employ a quantitative measure of the local rotation of gas, the circulation of a fluid $\Gamma$, which is defined as a line integral of the velocity field along a closed path and corresponds to a macroscopic measure of rotation. We define a two-component model for gas motions with a smooth function for large scales and a generating function $\mV(k)$ to model the noncircular motions. The velocity field arising from $\mV(k)$ behaves as a GRF. We compare the contributions from each component to the total measured circulation. In this framework, on galactic scales the contribution of noncircular motions to the circulation is negligible compared to the large-scale ordered rotation. On the smallest scales the circulation field is given mostly by $\mV(k)$. In other words, changes in the behavior of the observed distribution of $\Gamma$ at different scales illustrate how the dynamics transitions from circular to noncircular motions. With this in mind, we can define a spatial scale $\lequi$ at which large-scale rotation and noncircular motions contribute equally to the measured circulation of gas.

To test whether circulation is a useful tool to find the transition scale between galactic rotation and noncircular motions, we use hydrodynamical simulations of galactic disks with different initial conditions. Numerical simulations are an excellent test bed for the study of circulation since they provide the full velocity field and allow us to look for observable signature by changing different physical parameters, such as rotation or self-gravity. In future work we plan to expand these methods to make them applicable to high-resolution observations of gas velocity fields, like those being produced by the PHANGS project.

The paper is organized as follows. In section \ref{sec:methods} we introduce the main quantities analyzed in this work, vorticity and circulation, and we describe our technique to study the circulation in galaxies. In section \ref{sec:results} we describe the simulations and the application of our technique in those objects. We discuss our results in section \ref{sec:discussion} and list possible caveats of our simulations in section \ref{sec:Limitations}.

\section{Method}
\label{sec:methods}
\begin{figure}
\label{fig:vorticity_example}
\includegraphics[width=\linewidth]{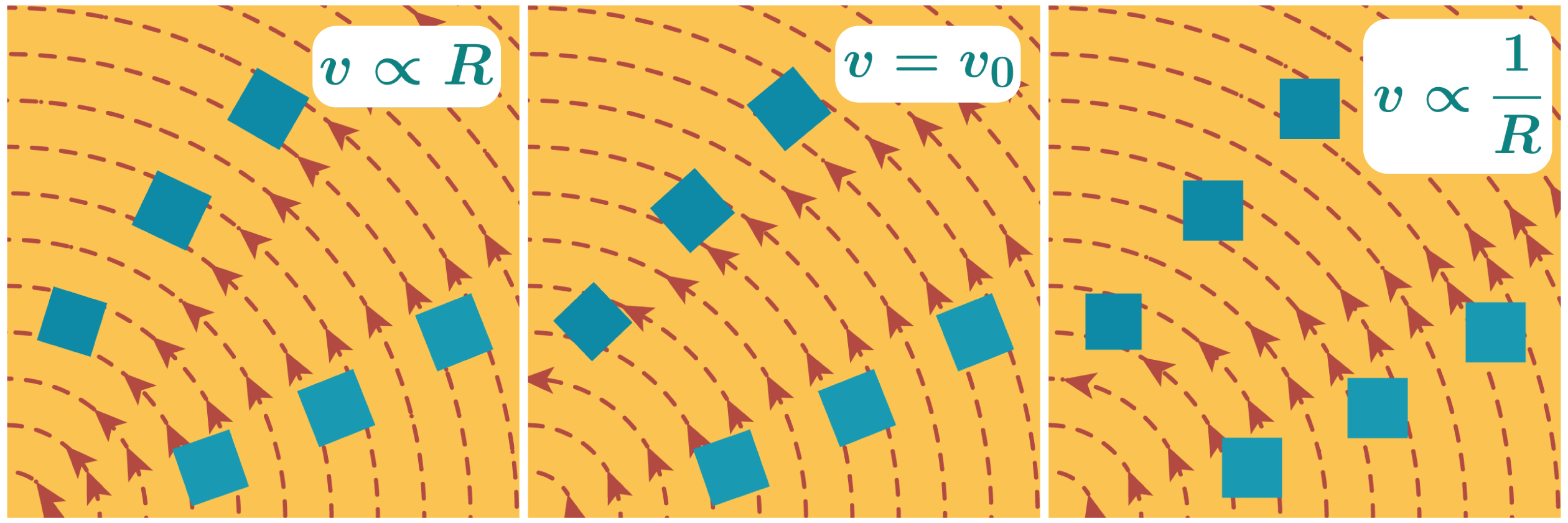}
\caption{ Illustration of the local rotation of fluid elements for different azimuthal velocity fields. Blue squares represent the fluid elements in two different positions in their path. The positions on each panel are the same and hence do not correspond to a specific instant in time. Left: solid-body rotation, $v \propto R $. The local rotation of any fluid element (blue squares) is equal to the galactic angular velocity $\Omega$. Middle: flat velocity curve, $v=v_0$. The local rotation of fluid elements $\Omega/2$. Right: irrotational flow, $v\propto 1/R$. For this flow the local rotation of a fluid element is zero. }
\end{figure}

\subsection{Vorticity and Circulation in Gas Dynamics}
\label{sec:termi}

One of the most useful notions in fluid dynamics is the vorticity vector, $\bomega$. In simple words, vorticity is a measure of the local rotation, and its direction is parallel to the spin of a fluid element. An infinitesimal fluid element experiences a rotation of $2\bomega$ respect to a local inertial reference frame. The vorticity is given by the curl of the velocity field
\begin{equation}
\label{eq:defvorticity}
\bomega=\nabla\times \bvel.
\end{equation}

To make this concept clearer, let us imagine a fluid with a circular velocity field $\bvel= v(R) \hat{\phi}$, where $\hat{\phi}$ is the azimuthal unit vector in cylindrical coordinates. In this scenario, the vorticity is 
\begin{equation}
\label{eq:vorticity}
\bomega= \frac{v(R)}{R}\left(1+\frac{\partial \ln v(R)}{\partial \ln R}\right)\hat{z}.
\end{equation}

We illustrate different velocity fields in Figure \ref{fig:vorticity_example}: solid-body rotation ($v\propto R$), a flat velocity curve ($v = v_0$), and an irrotational fluid ($v\propto 1/R$). We show three fluid elements in two arbitrary positions along their orbits. On each panel we show the fluid elements at the same azimuthal angle. The aim is to compare how much a fluid element has rotated once it passes by the same position on the disk.
In the case of solid-body rotation $\bomega=2\Omega \hat{z}$, where $\Omega(R)=v_0/R$, and patches of gas experience a local rotation of $\Omega$ with respect to a local inertial reference frame. For a flat velocity curve $\bomega=\Omega \hat{z}$, and the local rotation is half the galactic rotation $\Omega/2$. We can see in the middle panel of Figure \ref{fig:vorticity_example} that each fluid element has completed half the rotation of the left panel. We notice from equation \ref{eq:vorticity} that there is a critical case when $v(R) \propto 1/R$. For such a velocity field, $\bomega=0$ and fluid elements experience no local rotation with respect to an inertial reference frame. This kind of fluid is called irrotational, illustrated in the right panel of Figure \ref{fig:vorticity_example}.

There are two useful relations between vorticity and two quantities that are very helpful to have in mind. First, for a fluid with a circular velocity field $\bvel=v(R)\hat{\phi}$, the vorticity is proportional to the local angular momentum $\pmb{L} \propto \bomega$,  as demonstrated in Appendix \ref{sec:vorticityangularmomentum}. Second, for a circular velocity field the vorticity is related to $\Omega$ and $\kappa$ by $\bomega=\frac{\kappa^2}{2\Omega} \hat{z} $. This implies that for an irrotational fluid $\kappa=0$ and the Toomre parameter $Q=0$. Any perturbation larger than the thermal Jeans scale, $\lambda_{\rm J}\equiv c_s^2/\pi G \Sigma_{\rm gas}$, is gravitationally unstable. It is noteworthy that in an irrotational fluid its angular velocity $\Omega \neq 0$ and its shear $\frac{\partial \Omega}{\partial r} \neq 0$. 
Given the relation between vorticity and local angular momentum, this is not a surprising implication. A parcel with no angular momentum does not have rotational support to halt gravitational collapse.

Unfortunately, vorticity is a local quantity, defined for an infinitesimal fluid element. For finite regions of space it is better to compute the fluid circulation $\Gamma$, which corresponds to a macroscopic measure of rotation \citep{geo_book}. $\Gamma$ is defined as a line integral of the velocity field along a closed path,
\begin{equation}
\label{eq:line_integral}
\Gamma = \oint_{\delta S} \bvel\cdot d \pmb{l}. 
\end{equation}
For a continuous velocity field we can apply Stokes's theorem, which relates the line integral along a closed path to the surface integral over the area enclosed by it. This allows us to make the connection between $\Gamma$ and $\bomega$:
\begin{equation}
\label{eq:gamma_vort}
\Gamma = \oint_{\delta S} \bvel\cdot d \pmb{l} = \int_S \boldsymbol{\nabla} \times \bvel\ \cdot\ d\pmb{S}=\int_S \bomega \cdot d\pmb{S}.
\end{equation}

\begin{figure}[h]
\label{fig:circulation_example}
\includegraphics[width=\linewidth]{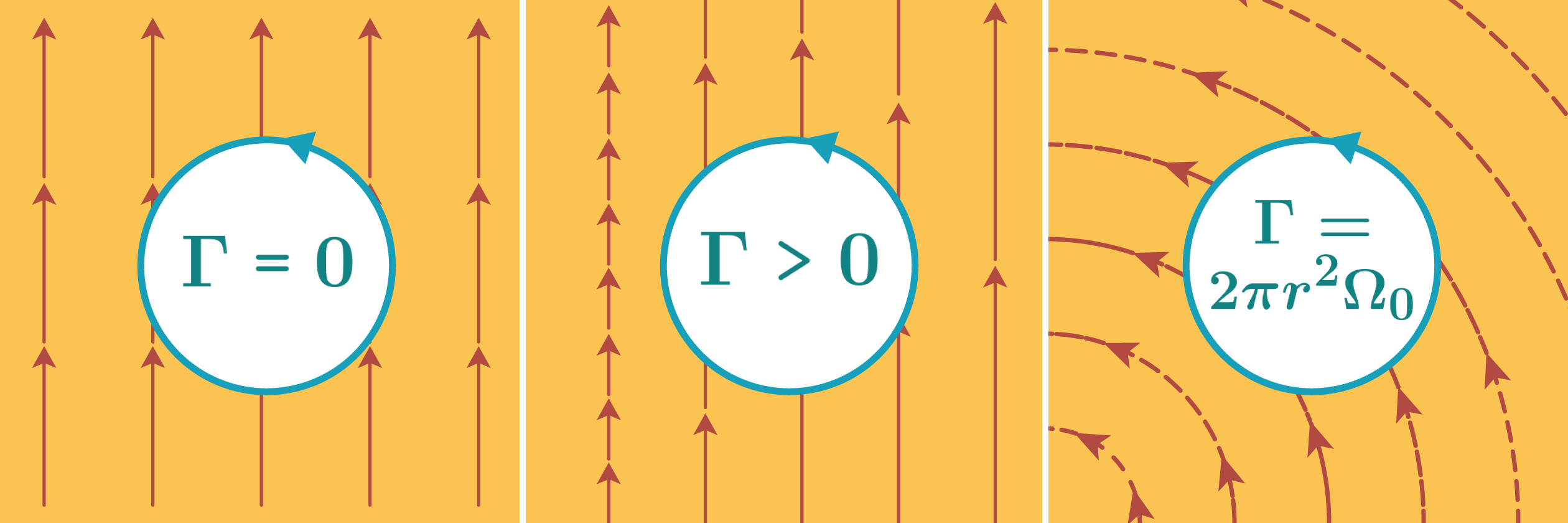}
\caption{ Illustration of the circulation in a closed region for different velocity fields. Left: constant velocity field. Middle: shear across the x-axis, $v_x=0$ and $v_y\propto x$. Right: solid-body rotation with angular velocity $\Omega_0$. The length of the arrows represents the magnitude of the velocity field. }
\end{figure}

We can think of circulation as the area-weighted integral of the vorticity field. 
Figure \ref{fig:circulation_example} illustrates the circulation around a circular path for different velocity fields. A fluid with a constant velocity $v_0$ has $\Gamma=0$, as it is constant in both magnitude and direction, and the line integral cancels out owing to the change of direction of the path with respect to the velocity field. In other words, bulk displacements make no contribution to $\Gamma$. A shear velocity field of the form $\bvel = (0,\Omega_0 x)$ has $\omega=\Omega_0$ and $\Gamma = \pi r^2 \Omega_0$. The last example shows solid-body rotation, $\bvel=(-\Omega_0 y, \Omega_0 x)$, with $\omega=2\Omega_0$ and $\Gamma=2\pi r^2\Omega_0$. 

However, realistic velocity fields are not completely smooth and behave differently at different scales. For instance, we might find that over a region $S$ the circulation is $\Gamma =0$. But this does not imply that $\omega=0$ over the whole region. Since the circulation is defined as a sum, if we divide $S$ into $N$ small subregions $s_i$, and $\Gamma = 0$, then

\begin{equation}
\Gamma_S=\int _S \bomega\cdot d\pmb{S} = \int _{s_1+ ... +s_N} \bomega\cdot d\pmb{S}= \sum_{i=1}^N \Gamma_{s_i} = 0. 
\end{equation}

The sum of all $\Gamma_i$ gives zero. There are infinite ways to distribute the values of $\Gamma_{i}$ to get $\Gamma=0$. The exact distribution of the circulation at this smaller scale will depend on the nature of the velocity field. This implies that only a multiscale measurement of the circulation can characterize the velocity field.

To have a full picture of the rotation of a fluid, we need to compute the circulation of gas at each point in the fluid on regions of different sizes. In this way we can create distributions of circulation at each spatial scale. To compare $\Gamma$ at different scales, let us define the normalized circulation $\gamma$:
\begin{equation}
\gamma = \dfrac{\int _S \bomega\cdot d\pmb{S}}{\int_S d{S}} = \dfrac{\Gamma}{A}, 
\end{equation}
where $A$ is the area of $S$. For solid-body rotation with angular velocity $\Omega_0$, $\gamma=2\Omega_0$ for any fluid patch, and the distribution of $\gamma$ will be a Dirac delta function $\delta(\omega-2\Omega_0)$. In the case of a rotating fluid with added random motions the distribution of $\gamma$ will be broader at small scales and will get narrower as we increase the size of the region in question, since we are adding random numbers and then dividing by a larger area. Hereafter and for simplicity we will refer to the {\it normalized circulation} simply as the {\it circulation} unless explicitly stated.

We can extend these ideas to the case of galactic dynamics. Imagine that the velocity field is composed by an ordered and smooth circular velocity field, $\bvel_{\rm rot}$, and a noncircular, random field, $\bvel_{\rm nc}$, i.e. $\bvel=\bvel_{\rm rot}+\bvel_{\rm nc}$. This gets translated into two components of the circulation field $\gamma= \grot +\gno$.  At large scales, $\gamma\approx\grot$, since the major contribution comes from galactic rotation and random or noncircular components cancel each other. At small scales, the distribution of $\gno$ gets broader, while the distribution of $\grot$ converges to $\orot$ given by equation \ref{eq:vorticity}. 

In brief, the probability density function (pdf) of $\gamma$ follows $pdf(\gamma) \simeq pdf(\grot)$ at galactic scales, and $pdf(\gamma) \simeq pdf(\gno)$ at small, parsec scales. This implies that at a particular scale
 $\lequi$, $\gno = \grot$, i.e. the velocity fields $\vrot$ and $\vno$ contribute equally to the measured circulation. Since $\lequi$ depends on the local properties of $\vrot$ and $\vno$, this scale has different values in different regions of a galaxy. For example, near the center of a galactic disk, $\orot$ is higher and the transition to non ordered motions occurs at smaller scales, i.e. smaller $\lequi$. In Section \ref{sec:compute_leq} we define how to compute $\lequi$. We analyze in detail the behavior of $\lequi$ for our set of simulations in Section $\ref{sec:discussion}$.

\subsection{The Method: Circulation as a Diagnostic of Gas Dynamics}

We model the velocity field as the sum of two fields with different properties: the first being an axisymmetric and smooth velocity field, $v_{\rm rot}=R\Omega(R)$, which is given by galactic rotation. The second field corresponds to a GRF, $\vno$. GRFs are fields that follow a Gaussian distribution. In this paper, we will use a continuous GRF that is defined by a generating function in Fourier space that specifies the contribution from each spatial scale to the random velocity field \citep{Lang_11}. These kinds of random fields are widely used in cosmology to model the primordial perturbations of the density field \citep{Pranav_19}. In $\vno$ we are including any source of noncircular large-scale motions, such as outflows, collapse, turbulence, and other large-scale coherent motions like those induced by spiral arms and bars. While many of these motions are not expected to be random or Gaussian, this is a good first-order approximation for a statistical description. In future work we can study better models for each component of the velocity field.

The aim is then to obtain the three fields, $\omega$, $\orot$, and $\ono$, from our simulations. We can compute $\omega$ directly from the simulations. To get $\orot$ we need to choose how to model the smooth profile of the velocity field (i.e. the rotation curve). Finally, we will model the random component by means of a function in Fourier space on the spatial coordinates. We have to point out that since we are computing the vorticity field for a discrete grid, this field is also $\gamma$ at the resolution level.

\subsubsection{Vorticity Field}
\label{sec:get_vort}
We calculate $\omega(x,y)$ as follows. First, we compute the two-dimensional velocity field averaging along the z-axis:
\begin{equation}
\displaystyle \pmb{V} (x,y) = \frac{\int_{-z_0}^{z_0} \bvel(x,y,z) \rho(x,y,z) dz}{\int_{-z_0}^{z_0} \rho(x,y,z) dz},
\end{equation}

where $\rho$ is the gas density, $\bvel=(v_x,v_y,v_z)$ is the three-dimensional velocity field, and $\pmb{V}=(V_x,V_y)$ is the reduced two-dimensional field. We choose $z_0= 1$ kpc over the whole galactic plane. 
Then $\omega(x,y)$ is given by
\begin{equation}
\omega(x,y) = \frac{\partial V_y(x,y)}{\partial x}  - \frac{\partial V_x(x,y)}{\partial y}.
\end{equation}

Note that we are considering all the gas in $z \in $ [-1kpc, 1kpc] to compute the integrated velocity fields, which is about 20 times the scale height of our simulated galaxies. We are not using a density threshold to integrate the velocity field. In observations, different tracers do not necessarily trace all the gas and are biased toward high-density regions.

\subsubsection{Smooth Component}
\label{sec:get_rot}
Since the definition of $\orot(R)$ involves radial derivatives, we choose to parameterize $\orot$ by an analytic function. To get $\orot$, we fit a rotation curve of the form
\begin{equation}
\label{eq:analytic}
V_{\rm rot}(R)=v_0 \arctan(R/R_1)\exp(-R/R_2)    
\end{equation}
to the circular velocity field, and we apply equation \ref{eq:vorticity} to obtain $\orot$. The $\arctan(x)$ function provides a good fit to observed rotation curves \citep{Courteau_97}, while the $\exp(-x)$ function recovers the decay of the rotation curve for our simulated galaxies. To fit the function in equation \ref{eq:analytic}, we divide the disk in radial bins of width 500 pc. For each radial bin we have a pair velocity uncertainty $(v_i,\delta v_i)$, where $v_i$ is the median of the circular velocity and $\delta v_i$ is half of the difference between the 84th and 16th percentiles of the circular velocity field. Finally, we perform a least-squares optimization to fit the rotation curve.
In Appendix \ref{app:rotation}, we show how our results change using a different model for the rotation curve, e.g. adding of a more sophisticated measure of the large-scale motions via harmonic decomposition. We find that our results are not very sensitive to the choice of the model of galactic rotation.

\subsubsection{Random Component}
\label{sec:random_component}
The final step is choosing a model for $\vno$ to obtain its contribution to the vorticity field. Here we choose $\vno$ to be defined by a generating function $\mV(k)$ in Fourier space. The relation between $\mV(k)$ and the field $\vno$ is derived in Appendix \ref{sec:random_fields}. For a two-dimensional field, $\mV(k)$ is related to the energy power spectrum $E(k)$ by
\begin{equation}
\dfrac{1}{2}\langle \vno^2 \rangle = \pi \int \mV(k)^2 k dk = \int E(k)dk.
\end{equation}
This relation is shown in equation \ref{eq:sigma_0}. The function $\mV(k)$ is not unique for a whole galaxy. Each region of a galaxy is subjected to different conditions of stability, feedback, and dynamics, which will give rise to different noncircular velocity fields (turbulence, collapse, and gas flows). Given the geometry of disk galaxies, we expect that the dynamics of gas change across galactocentric radius. The simplest way to approach these differences is to separate the galaxy into radial bins. Then, each radial bin will be described by its own function $\mV(k)$. 

We choose $\mV(k)$ to be of the form
\begin{equation}
\mV(k)\propto
\begin{cases}
 & \ k^{-n_1}\quad \text{ if } \kmin< k< \kc\\ 
 & \ k^{-n_2}\quad \text{ if } \kc \leq k < \kmax ,\\ 
 & 0 \quad \qquad \text{elsewhere} \\
\end{cases}
\end{equation}

where the wavenumber k is related to the spatial scale as $k=\lambda ^{-1}$ (no $2\pi$ factor). We fix $\kmin$ and $\kmax$ to $4/L$ and $1/(4 \Delta x)$, respectively, where $L$ is the box size and $\Delta x$ is the spatial resolution of the two-dimensional field, 30 pc for the simulations described in Section \ref{sec:simulations}. This limits the dynamic range of $\mV(k)$ between 120 pc and 10 kpc (choosing $L=40$ kpc). If we add the constraint of continuity for $\mV(k)$ at $k_c$, we need a parameter to set the amplitude of $\mV(k)$ which translates into the amplitude of the velocity field. We choose this last parameter to be the characteristic velocity dispersion of the random velocity field, given by
\begin{equation}
\label{eq:sigma0_1}
\sigma_0 ^2 =  2 \pi \int_{k_{\rm min}}^{k_{\rm max}} \mV(k)^2 k dk.
\end{equation}

This is a property of GRFs \citep{Lang_11}. To resume, the parameters defining the velocity field are ($n_1$, $n_2$, $\kc$, $\sigma_0$). The vorticity of this velocity field can be calculated using equation \ref{eq:defvorticity} or directly from the parameters using equation \ref{eq:omega_model}. We denominate the velocity and vorticity fields obtained from the $\mV(k)$ as $\vno$ and $\ono$.  In summary, our model for the vorticity field is
\begin{equation}
\label{eq:model}
\omega(x,y) = \omega_{\rm rot}(R)+\ono(n_1(R),n_2(R),\kc(R),\sigma_0(R)).   
\end{equation}

We test two additional models for $\mV(k)$ with $n_1=n_2$ i.e. a single power law, shown in Appendix \ref{app:power}. In one model $\kmax$ is fixed to $(4\Delta x)^{-1}$ while in the second we allow $\kmax$ to vary. A single power law with fixed $\kmax$ that best fits the noncircular component of $\gamma$ does not match the behavior of $\gno$ as a function of scale. This implies that is not a good representation of $\gno$.  On the other hand, a single power law with variable $\kmax$ gets similar results compared to the piecewise power law used in this work. This is consistent with the possible values that we find for $n_2$ in Section \ref{sec:distributions} (i.e. it mimics the adopted model with a high value of $n_2$).

\subsubsection{Distribution of $\gamma$}

To find the best values for $n_1$, $n_2$, $\kc$ and $\sigma_0$, we measure the distribution of the circulation $\gamma (x,y,\ell)$ 
\begin{equation}
\displaystyle \gamma(x,y,\ell) = \frac{1}{\ell ^2} \int \omega (x,y) dS    
\end{equation}

at each scale $\ell$, integrating over square regions of size $\ell \times \ell$ centered on points $(x,y)$. According to our model, $\gamma = \grot+\gno(n_1n,n_2,\lk,\sigma_0)$. Since $\vno$ is a GRF, $\ono$ and $\gno$ are GRFs too. In equation \ref{eq:sigma_equation} we show that the variance of $\gno(x,y,\ell)$ is determined by $\ell$ and the generating function $\mV(k)$. Then, we only need to compute the variance of $\gno$ as a function of $\ell$ and use equation \ref{eq:sigma_equation} to find the parameters of $\mV(k)$.

\subsubsection{The Scale at Which Gas Noncircular Motions Start to Dominate}
\label{sec:compute_leq}
The last and most relevant quantity in our framework is $\lequi$, the scale at which the contributions from $\grot$ and $\gno$ to the measured circulation of gas are roughly the same. For $\ell>\lequi$ the circulation of gas is dominated by galactic rotation, while at scales $\ell<\lequi$ it is dominated by noncircular motions.

We define the scale $\lequi$ in the following way. At each scale $\ell$ we measure the ratio $ f_{\gamma}(\ell)=\sum  \gamma_{\rm nc, \ell} ^2 / \sum  \gamma_{\rm rot, \ell}^2$. Then, $\lequi$ is given by the equation 
\begin{equation}
f_{\gamma}(\lequi)=1.0  \quad \text{with} \quad  f_{\gamma}(\ell)=\dfrac{\sum  \gamma_{\rm nc, \ell} ^2 }{ \sum  \gamma_{\rm rot, \ell}^2}.
\end{equation}

We compare their squared values since $\gamma$ can have negative values. 
For a random variable $x$ with mean value $\mu_x$ and standard deviation $\sigma_x$,  the expected value of $\sum x^2$ is $\mu_x^2 +\sigma_x ^2$. We can rewrite $f_\gamma(\ell)$ as 
\begin{equation}
f_\gamma(\ell)=\frac{\mu_{\gno}^2(\ell) +\sigma_{\gno}^2(\ell)}{\mu_{\grot}^2(\ell) +\sigma_{\grot}^2(\ell)}
=\frac{\sigma_{\gno}^2(\ell)}{\mu_{\grot}^2(\ell) +\sigma_{\grot}^2(\ell)},
\end{equation}
where $\mu_{\gno}$, $\mu_{\grot}$, $\sigma_{\gno}$, and $\sigma_{\grot}$, are the mean values and standard deviations of $\gno$ and $\grot$ as a function of $\ell$. 
For a region with constant $\omega_{\rm rot}$,  $f_{\gamma}(\lequi)=1.0$ is equivalent to the equation $\sigma_{\gno}(\ell=\lequi) = \omega_{\rm rot}$. Given the parameters that define $\grot$ and $\gno$, $\lequi$ is a function of the form $\lequi(\omega_{\rm rot},n_1,n_2,\lk,\sigma_0)$.\\

\subsubsection{Deriving $\gno$ and $\lequi$ }

In practice, we are looking for the parameters $n_1$, $n_2$, $k_c$, and $\sigma_0$, that best represent the equation $\gamma=\grot+\gno$. However, for a given set of parameters, the field $\gno$ is a random realization from a parent GRF and is not single valued. This means that the expression $\gamma=\grot+\gno$ has to be considered as the sum of two distributions rather than the sum of two fields or images. Then, to look for parameters that can model our data, we need to compare $\gamma$, $\grot$, and $\gno$ as distributions. We choose to compare histograms of $\gamma$ and $\grot+\gno$  as the final ingredient of our technique.

Our method can be summarized as follows:
\begin{itemize}
\item[1.]Measure the two-dimensional vorticity field $\omega(x,y)$.
\item[2.]Model the large-scale and axisymmetric component of the velocity field, $V_{\rm rot}$ using equation \ref{eq:analytic} and compute $\omega_{\rm rot}$.
\item[3.]Divide the disk into different radial annuli.
\item[4.]From $\omega$ and $\omega_{\rm rot}$, measure the distributions of $\gamma$ and $\grot$ at each scale $\ell$ within each radial annulus.
\item[5.]At each spatial scale we compare the distributions of $\gamma$ and $\grot + \mathcal{N}(0,\sigma_{\gamma})$, where $\mathcal{N}(0,\sigma_{\gamma})$ is a random field with dispersion $\sigma_{\gamma}$. We fit $\sigma_{\gamma}$ using the least-squares method. This step creates an array $\sigma_{\gamma}(\ell)$ as a function of $\ell$ with its respective uncertainty $\epsilon_{\sigma}$.
\item[6.]Explore the parameter space $\btheta=(n_1,n_2,k_c,\sigma_0)$ of the function $\mV(k)$. Each parameter vector $\btheta$ defines a different curve $\sigma_{\gno}(\ell,\btheta)$. To find the posterior distributions of $\btheta$ given our data $\mathbf{D}$, we use the Bayes's theorem: 
\begin{equation}
\label{eq:Bayes}
P(\btheta | \mathbf{D}) =\dfrac{P(\mathbf{D} | \btheta) P(\btheta)  }{P(\mathbf{D})} 
\end{equation}
In this equation, $P(\mathbf{D} | \btheta)$ is the likelihood to obtain $\mathbf{D}$ that in our case corresponds to the array $\sigma_{\gamma}(\ell)$. Our likelihood is given by
\begin{equation}
\label{eq:likelihood}
P(\mathbf{D} | \btheta) = \prod _{\ell} \dfrac{1}{\sqrt{2\pi}\epsilon_{\sigma}} \exp\left(\dfrac{-(\sigma_{\gamma}(\ell)-\sigma_{\gno}(\ell,\btheta))^2}{2\epsilon_{\sigma}^2}\right)
\end{equation}
$P(\btheta)$ is our prior knowledge of the parameters. We assume uniform prior distributions for each parameter. $P(\mathbf{D})$ is the Bayesian evidence of the data that ensures proper normalization. To sample the posterior distributions, we use Markov Chain Monte Carlo methods. We use 72 random walkers that are updated using the Metropolis-Hastings algorithm. Each random walker creates a chain with 15,000 values of $\btheta$. For our analysis we ignore the first 5000 steps. This step creates samples of $\btheta =(n_1,n_2,k_c,\sigma_0)$, from which we reconstruct the pdf's of the model parameters. This sample also establishes the parent distribution of $\gno$. 
\item[7.] From $\grot$ and the distributions of $\gno$ we derive the distribution of the scale $\lequi$ at each radial annuli.
\end{itemize}

\begin{figure*}
\label{fig:density_maps}
\includegraphics[width=\textwidth]{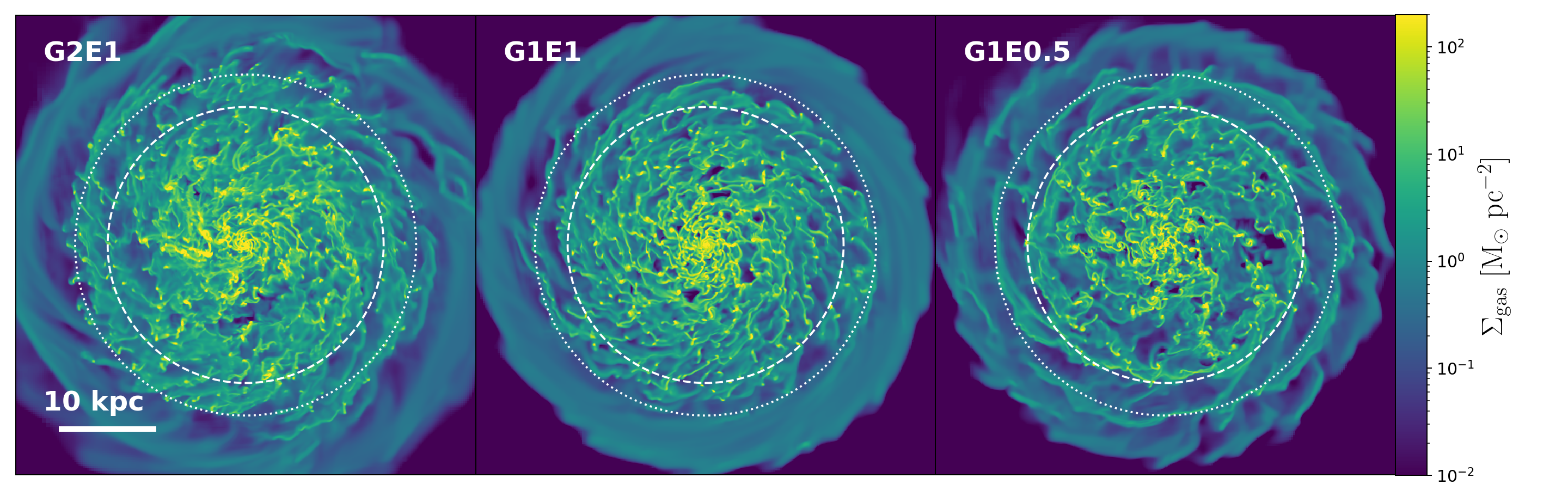}
\caption{Gas surface densities integrated along the z-axis at $t=700$ Myr. Dashed white circles delimit the region inside 15 kpc, which corresponds to the maximum radius of the defined annuli. The dotted line shows the maximum radius of the cells that are included in the analysis of circulation, which corresponds to 18.54 kpc. }
\end{figure*}

\subsubsection{Choice of Spatial Bin Size and Estimation of Uncertainties}

To build the histograms and to compare them, we must choose the width of the bins, $\Delta \gamma$, and the uncertainty, $\epsilon_{\gamma}$, for each measurement with their propagation in the histogram bins. For the bin width we choose a conservative criterion: 8 times the bin width set by the Freedman-Diaconis rule \citep{Freedman}, $\Delta \gamma = 16\ {\rm IQR}\ (\gamma)\ N^{-1/3}$, where $\rm IQR(\gamma)$ is the interquartile range of $\gamma$ and $N$ is the total number of data points. The Freedman-Diaconis rule attempts to minimize the integrated mean squared difference of the histogram model and the true underlying density. 

For the uncertainty in $\gamma$ we set an uncertainty of $\epsilon_v$ = 1 km s$^{-1}$ in the measured velocity at the resolution of the simulation $\Delta x$ ($\approx 30$ pc), which is comparable with the precision of recent gas velocity measurements on nearby galaxies \citep{Druard_14,Caldu_16,Koch_18, Sun_2018}. 
To propagate the uncertainty, we use equation \ref{eq:line_integral}. For a square region with area  $\Delta x \times \Delta x$, i.e. at the maximum resolution, the circulation is the sum of the integral along the four faces of the square. Then, $\Gamma = v_1 \Delta x + v_2  \Delta x +v_3 \Delta x +v_4 \Delta x$ (with the respective signs due to the dot product) and its uncertainty is $\epsilon_{\Gamma}=\sqrt{4 \epsilon_v^2}  \Delta x = 2 \epsilon_v \Delta x$. For $\gamma$ the uncertainty is $\epsilon_{\gamma}=2\epsilon_v/\Delta x$.
A square region of size $\ell=N \Delta x$ is delimited by $4N$ linear segments, $N$ at each side. Then, $\Gamma=\sum_{i=1}^{4N} v_i \Delta x$ and its uncertainty is $ 2\sqrt{N}\epsilon_v\Delta x$, while for $\gamma$ it is $2\sqrt{N}\epsilon_v\Delta x / (N \Delta x)^2 = (2\epsilon_v/\Delta x) \times N^{-3/2} = \epsilon_{\gamma} N^{-3/2}$. The scaling of the uncertainty of $\gamma$ with $N$ is not straightforward if we use equation \ref{eq:gamma_vort}. We have to recall that $\omega$ is the difference between two terms. If we add the vorticity of two neighbor cells, we are also subtracting the line integral along the line that both regions share. For that reason, when we compute the uncertainty in $\gamma$ for a region with $N\times N$ elements, we are adding $4N$ terms instead of $N\times N$.

The histogram counts of our distributions have Poisson noise, i.e. an uncertainty of $\sqrt{N}$, where $N$ is the number of data points lying in a given bin. We also have to propagate the uncertainties in $\epsilon_{\gamma}$ into the histogram. Let us consider the $j$th bin with endpoints [$l_j$,$u_j$] and a data point with value $\gamma_i$ and uncertainty $\epsilon_i$. Assuming that $\gamma_i$ is the mean value of a Gaussian random variable with standard deviation $\epsilon_i$, the probability $p_{ij}$ that this data point lies within [$l_j$,$u_j$] is 
\begin{equation}
\displaystyle p_{ij} = \frac{1}{2} \left[\erf\left(\frac{u_j - \gamma_i}{\sqrt{2}\epsilon_i}\right) -\erf\left(\frac{l_j - \gamma_i}{\sqrt{2}\epsilon_i}\right)\right]
\end{equation}

where $\rm erf$ is the error function. Each bin acts like a Bernoulli random variable: we add 1 if the measurement lies in the bin or zero otherwise, with a probability $p_{ij}$ and $1-p_{ij}$, respectively. The variance for the Bernoulli distribution is $p_{ij}(1-p_{ij})$. Then, the total variance in the $j$th bin due to the uncertainties in $\gamma$ is $\sum_{i}^{N} p_{ij}(1-p_{ij})$.

\begin{figure}
\label{fig:rotation_curve}
\includegraphics[width=\linewidth]{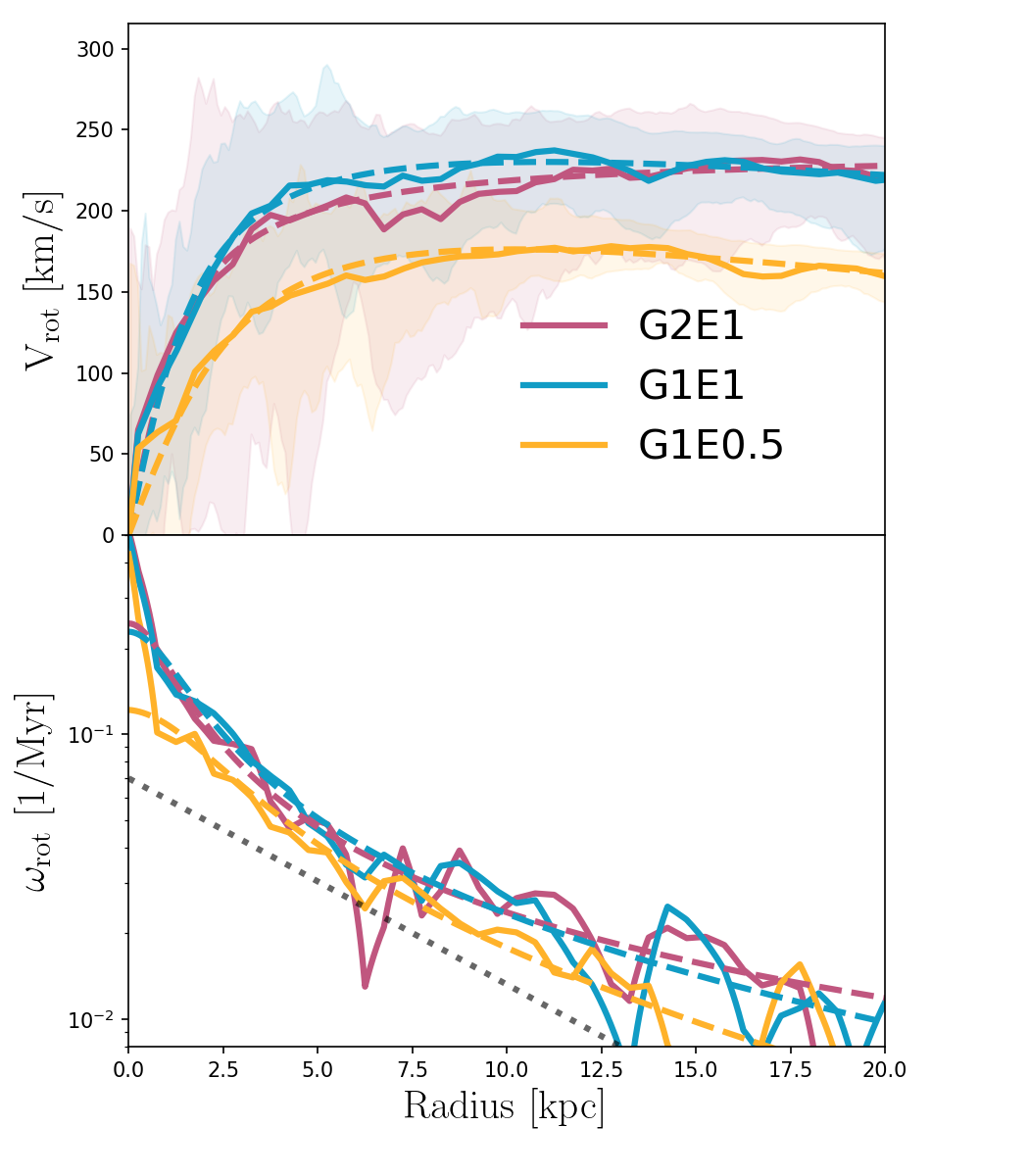}
\caption{Profiles of rotational velocities and large-scale vorticity. Top: radial profiles of the circular velocity $v_{\rm rot}$, which are used to compute the large-scale vorticity component $\omega_{\rm rot}$. Solid lines show the median values of the rotation curve. Shaded regions represent 1$\sigma$ uncertainties due to variations in the velocity field. The analytic models of the rotation curves are shown as the dashed lines. Bottom: radial profiles of vorticity $\omega_{\rm rot}(R)$. The solid and dashed lines correspond to the median values and the analytic models, respectively. The gray dotted line shows a function proportional to $\exp (-R/5 {\rm kpc})$ for comparison.}
\end{figure}

\subsection{Parameter Distributions}

To sample the distributions for $n_1$, $n_2$, $k_c$, and $\sigma_0$, we use the Metropolis-Hastings algorithm to generate Markov chains using equation \ref{eq:Bayes}. To compute the likelihood function in equation \ref{eq:likelihood}, we need to calculate the integral in equation \ref{eq:sigma_equation} each time we test a new set of parameters. Since this step is computationally expensive, we divide the subspace ($n_1$, $n_2$, $k_c$) into a $72^3$ grid over which we pre-tabulate the integral. The parameter $\sigma_0$ works as a normalization of $\mV(k)$ and can be handled independently. The intervals chosen are $n_1 \in (0.8,2.5)$, $n_2 \in (2.0, 20.0)$ and $\kc \in (2\kmin,\kmax/2)$. 

\section{Application to Simulated Galaxies}
\label{sec:results}

We start this Section by introducing the simulations used as a test bed for our method. 

\subsection{Simulations}
\label{sec:simulations}

We test our technique on three hydrodynamical simulations of disk galaxies. In each simulation we apply our circulation-based method on nine radial annuli, 3 kpc wide, centered on 1.5, 3.0, 4.5, 6.0, 7.5, 9.0, 10.5, 12.0, and 13.5 kpc. We vary the scale $\ell$ from the maximum resolution of the simulations 30 pc-5 kpc. Notice that $\ell$ can be larger than the width of a radial annulus. We might argue that within a radial annulus there is no information of scales larger than the width of the radial bin. However, the function $\mV(k)$ has information about the correlation between two points in the velocity field, and within each radial annulus we can find points separated by distances larger than 3 kpc. The center of each square region is inside the 3 kpc annulus, but it can cover cells outside the annulus. Information from neighbor regions will affect the values of the parameters within each annular region. This might smooth the resulting radial profiles of our model parameters.

To create this set of simulated galaxies, we use the adaptive mesh refinement code Enzo \citep{Bryan_14}. We run simulations of spiral galaxies for 700 Myr with a coarse resolution of 3.7 kpc. We use two criteria to refine a given gas cell, and both of which have to be fulfilled: refinement by baryon mass, and Jeans length. The Jeans length is at least resolved by four cells to prevent artificial fragmentation \citep{Truelove_97}. During the first 500 Myr, we use a maximum resolution of 60 pc until a quasi-steady state is reached. Then, the resolution is increased to 30 pc for 200 Myr. Over 80\% of the mass in gas cells is found at the highest resolution. To obtain images and the velocity fields of the simulated galaxies, we use the {\it yt} python package \footnote{yt : http://yt-project.org}\citep{Turk_11}.

These simulations are modeled as a four-component system that includes gas, star particles, and time-independent stellar and dark matter potentials. The time-independent stellar potential is given by the Miyamoto-Nagai profile \citep{Miyamoto_Nagai_75}, with a gravitational potential $\Phi_{\rm stellar} $ given by 
\begin{equation}
\Phi_{\rm stellar} = - \frac{GM_{\rm stellar}}{\sqrt{R^2+(a+\sqrt{z^2+b^2})^2}},
\end{equation} 

where $M_{\rm stellar}$ is the total stellar mass of the field and $a$ and $b$ are characteristic length scales. We choose $a=5$ kpc and $b=200$ pc, which are similar to the fitted values for the Milky Way stellar disk \citep{Kafle_14}. For the DM potential we use the Navarro-Frenk-White profile \citep{NFW97}. The virial mass of the DM potential and the total stellar mass of the stellar potential are different for each simulation, which will be defined later. Note here that the external potential is axisymmetric, and we do not include spiral structure as done in other works \citep[e.g.][]{Dobbs_15}. The gaseous disks start with an exponential radial profile with a radial scale length of $3.5$ kpc. 

Star particles are formed when (i) the local number density $n_{\rm cell}$ is greater than $\rm 100\ cm^{-3}$, (ii) the velocity field is converging $\nabla\cdot \bvel <0$, (iii) the cooling time is shorter than the local freefall time, and (iv) the gas mass in the cell $m_{\rm cell}$ is greater than the Jeans mass, $m_{\rm cell}>m_{\rm J}=\dfrac{4\pi}{3}(\lj/2)^3$. If these criteria are satisfied, a star particle is formed with a mass equal to $0.1\ m_{\rm cell}$. The typical mass of a star particle is of the order of 1$\times 10^4$ $\msun$ and represents a population of stars. Our simulations match the Kennicutt-Schmidt relation \citep{Kennicutt_98}. 

We include stellar feedback from supernovae (SNe), H II regions, and momentum injection from massive stars. To compute the energy or momentum injection as a function of time, for each type of stellar feedback we use tabulated results from STARBURST99 \citep{STARBURST99} assuming a Kroupa initial mass function, solar metallicity, and instantaneous star formation. We model SN feedback by injecting $10^{51}$ erg of thermal energy per every 55 $\msun$ of stars formed. This is 1.6 times more energy than the simulations of the Agora Project \citep{Agora2}. Since star particles represent a population of stars, the energy is deposited continuously at the cell where the star particle lies. To add the effects of H II regions, we follow the approach of \cite{Renaud_13} and \cite{Goldbaum_2016}, heating the gas up to $10^4$ K within the Str\"omgren radius. If the volume of the Str\"omgren sphere, $V_S$, is smaller than the cell volume, $V$, only a fraction $V_S/V$ of the thermal energy is deposited in the corresponding cell. In our simulations, most of the time $V_S<V$, and on average the gas is heated up to 7000 K. To compute the momentum injection, we consider stellar winds and radiation pressure \citep{Agertz_13}. To account for the radiation pressure, we compute the bolometric luminosity $L_{\rm bol}$ for each active stellar particle, and we distribute the momentum $L_{\rm bol}/c$ evenly in the six nearest cells. For stellar particles separated by one cell this causes some cancellation of the injected momentum \citep{Hopkins_Grudic}. We underestimate the effect of radiation since we do not consider the scattering of IR photons. For our simulations, and assuming an IR opacity if $\kappa_{\rm IR}=10\ {\rm cm^2g^{-1}}$, the optical depth of IR radiation is usually $\tau_{\rm IR}\simeq 0.2$. To model the energy lost by radiation, we use the cooling curves of \cite{Sarazin_87} for temperatures $T \geq 10^4$ K and those of \cite{Rosen_95} for $300 {\rm K} < T<10^4$ K. This imposes a minimum value for the Jeans scale $\lj$. For a surface density of $10 \msun/pc^2$ and a temperature $T=300$ K the $\lj$ is of the order of 100 pc. This means that overdensities in our simulations are more representative of H I clouds. \\

We run a second set of simulations using only SN feedback to explore the effects of changing the feedback prescription. This second set has a higher density threshold of $\rm 2800\ cm^{-3}$, necessary to match the Kennicutt-Schmidt relation \citep{Kennicutt_98}. Except from the expected decrease in the magnitude of noncircular motions due to the injection of less energy on small scales, most of the conclusions from the previous set of simulations hold also for these simulations, so they are presented and discussed in Appendix \ref{app:simulations}.

The three simulations discussed here are designed as follows: run G2E1 corresponds to our reference simulation, with $2\times 10^{10}$ $\msun $ of gas, a stellar potential of $10^{11}$ $\msun$ and a halo mass $M_{\rm 200}= 8\times 10^{11}$ $\msun$. The masses for the stellar and DM component are similar to Milky Way values \citep{Kafle_14}.  Run G1E1 is identical to the reference run in all parameters, except that the gas mass is reduced by half (i.e. it has a lower gas fraction). Run G1E0.5 has the same gas fraction and disk scale length as the reference run, but half the mass in all components (stars, gas, and DM), therefore being a low surface density version of the reference run. For G2E1 and G1E1 the concentration parameter of the DM potential is set to $c=21$, and for G1E0.5 it is tuned to maintain the same shape of the rotation curve, although the normalization can be different. The relevant parameters for these simulations are shown in Table \ref{table:simulation_parameters}. In our nomenclature G stands for the amount of gas and E for the magnitude of the external potential. We have to point out that these models sit a factor of three above the stellar mass-halo mass relation. These add changes in the magnitude of vertical acceleration, but we do not expect to produce major changes in the two-dimensional velocity field.

In Figure \ref{fig:density_maps} we show projections of the gas density field across the z-axis. These galaxies do not present grand-design spiral patterns. Near the galactic center we see clumps with different sizes for each simulation, particularly larger clumps for G2E1 that show signs of tidal interactions. Structures in G1E0.5 appear to be less affected by shear owing to the lower magnitude of its rotation curve. The dashed white circle of 15 kpc radius in Figure \ref{fig:density_maps} shows the outer edge of the outermost radial annulus where we measure the circulation of gas. When we measure the circulation within square regions of size 5 kpc, and centered on a point at radius 15 kpc, we include points that are up to a distance of 18.54 kpc from the galactic center. We show a 18.54 kpc radius dotted circle for illustrative purposes in Figure \ref{fig:density_maps}.

\begin{table}[ht]
\caption{Simulation parameters}
\centering
\begin{tabular}{c c c c}
\hline\hline
Run & $M_{\rm gas}$ & $M_{\star}$ & $M_{\rm 200}$ \\ [0.5ex] 
\hline
 & $(M_{ \odot})$ & $(M_{\odot})$ & $(M_{\odot})$\\
\hline
G2E1	    &	$2\times 10^{10}$	&	$1\times 10^{11}$	&	$8\times 10^{11}$\\
G1E1	    &	$1\times 10^{10}$	&	$1\times 10^{11}$	&	$8\times 10^{11}$\\
G1E0.5  	&	$1\times 10^{10}$	&	$5\times 10^{10}$	&	$4\times 10^{11}$\\

\hline
\end{tabular}
\label{table:simulation_parameters}
\end{table}

\subsection{Galactic Rotation}

We show the rotation curves $v_{\rm rot}(R)$ of each run in the top panel of Figure \ref{fig:rotation_curve}. The shaded regions show the variations in the tangential velocity field. The analytic rotation curves, obtained by fitting equation \ref{eq:analytic}, are shown as dashed lines. In the bottom panel of Figure \ref{fig:rotation_curve} we also show the resulting vorticity $\omega_{\rm rot}(R)$. The vorticity coming from galactic rotation, $\omega_{\rm rot}(R)$, decreases as we get far from the galactic center. This means that $\grot$ decreases with galactocentric radius. If the parameters of $\mV(k)$, which define the magnitude of $\gno$, were constant across galactocentric radius, the relative contribution from $\grot$ to the total circulation $\gamma$ would decrease with galactocentric radius, and the spatial scale $\lequi$ should increase with radius.

\begin{figure*}
\label{fig:parameters}
\includegraphics[width=\linewidth]{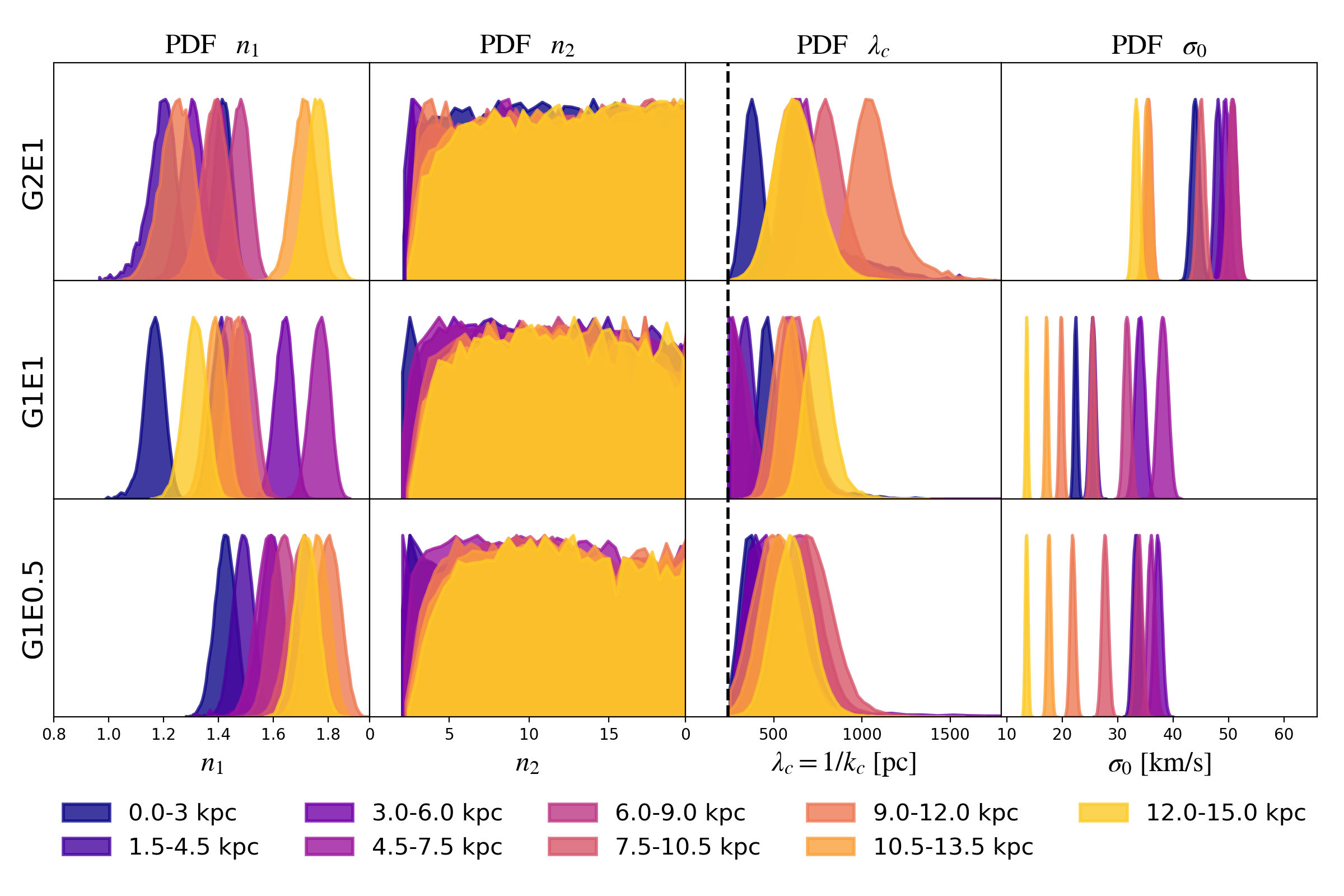}
\caption{Posterior distributions of model parameters. From top to bottom we plot the distributions for each of the three simulations. From left to right we show normalized distributions of $n_1$, $n_2$, $\lk=1/\kc$, and $\sigma_0$. The distribution for each annulus is shown with a characteristic color, going from purple to yellow with increasing galactocentric radius. The vertical black dashed line in the distributions for $\lk$ corresponds to eight times the resolution of the simulations, the minimum allowed value for $\lk$.}
\end{figure*}

\begin{figure*}[!ht]
\label{fig:corner}
\includegraphics[width=0.95\linewidth]{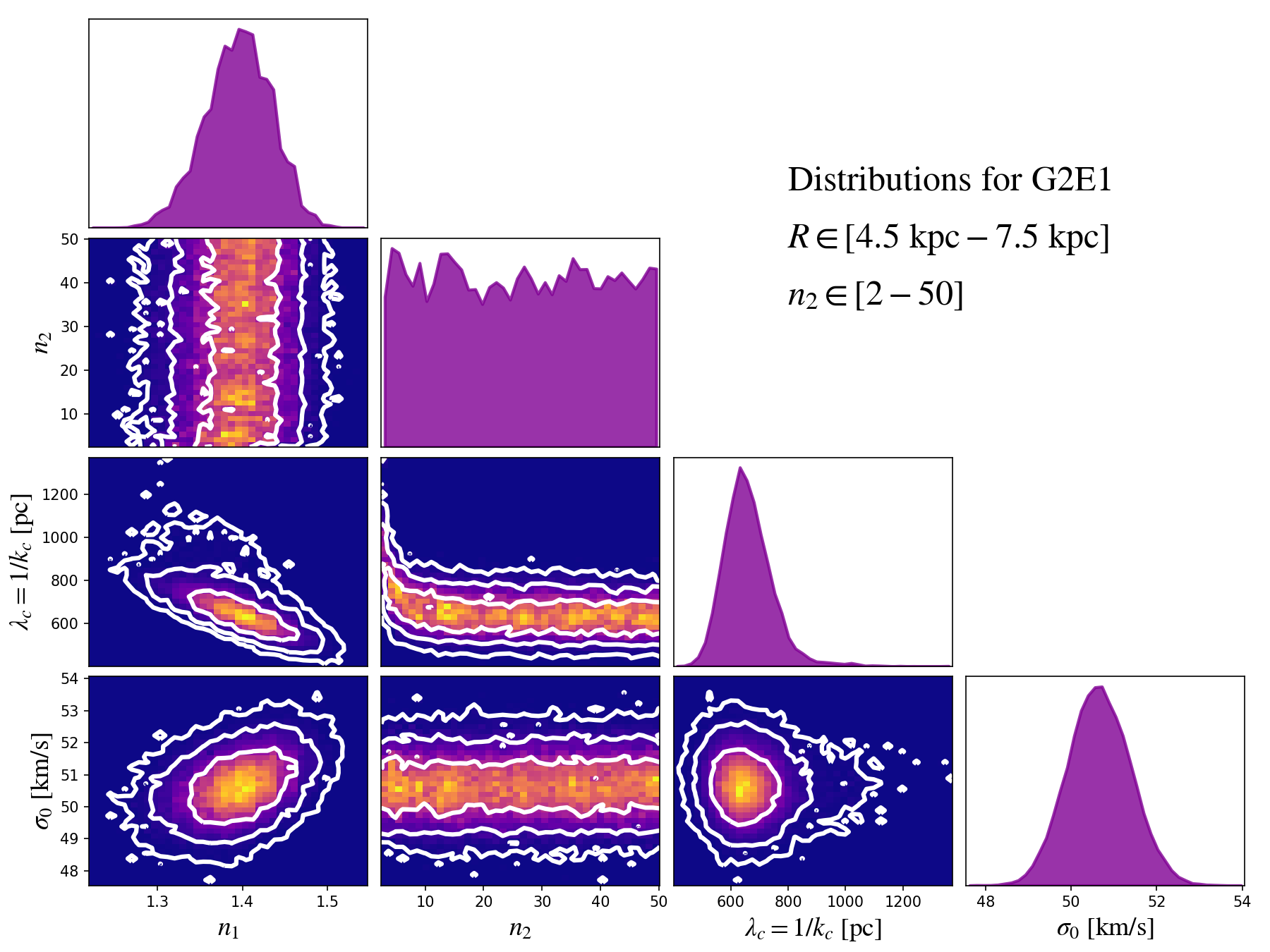}
\caption{Distributions of parameters $n_1$, $n_2$, $\lk=1/\kc$, and $\sigma_0$ for run G2E1, in the radial annulus [4.5 kpc - 7.5 kpc]. Purple histograms in the diagonal show the marginal posterior distributions for $n_1$, $n_2$, $\lk=1/\kc$, and $\sigma_0$ in descending order. Off-diagonal plots show two-dimensional histograms of the model parameters. White contours show the 68\%, 95\%, and 99.7\% confidence intervals. In these plots the explored range of values for $n_2$ has been extended to $n_2=50$.}
\end{figure*}

\subsection{Characterization of Random Motions}
\label{sec:distributions}
To analyze our simulations, we choose a box size of $L = 40$ kpc. This implies that $k_c$ is bounded between values $(2\kmin = 40\ {\rm kpc}/8)^{-1}=(5\ {\rm kpc})^{-1}$ and $\frac{1}{2}\kmax = (8\times 30\ {\rm pc})^{-1}=(240\ {\rm pc} )^{-1}$. In Figure \ref{fig:parameters} we show the probability distributions of $n_1$, $n_2$, $\lk$, and $\sigma_0$, where $\lk=1/k_c$ corresponds to the characteristic scale of the model $\mV(k)$. Keep in mind that, given the shape of the chosen function $\mV(k)$, the parameters $n_1$, $n_2$, and $\lk$ are correlated parameters. Runs G2E1 and G1E1 show variations of $n_1$ between 1.0 and 1.9. The distribution of $\lk$ ranges from its minimum value of $\sim$ 240 pc to 1.1 kpc. In the annulus of G1E1 centered at 4.5 kpc $\lk$ is unresolved. The parameter $\sigma_0$ shows narrow distributions. G2E1, the most massive galaxy, shows values of $\sigma_0$ above 30 km s$
^{-1}$ at all radii. The distribution of $n_2$ gets flat for values over 5. The model is not sensitive to variations of $n_2$ over $n_2=5$. Equation \ref{eq:sigma_vort_omega} shows the contribution from each scale and the amplitude of the vorticity field through an integral of $\mV(k)$. As the values of $n_2$ increase, the contribution from $k>k_c$ to the vorticity field starts to get smaller. This suggests that the function $\mV(k)$ can be approximated as a single power law with a cut at $k_c$, i.e. $\mV(k>k_c)=0$ as discussed in \ref{sec:random_component}. We show in Figure \ref{fig:corner} detailed distributions for G2E1. The off-diagonal plots of Figure \ref{fig:corner} show two-dimensional histograms of the model parameters that help to visualize the correlation between these parameters. We can see a high correlation between $\lk$ and $n_1$ and also a correlation between $\lk$ and $n_2$ for low values of $n_2$. The explored range of values for $n_2$ in Figure \ref{fig:corner} is extended to $n_2=50$ to show that its distribution remains uniform beyond $n_2=20$. For the annular regions, 1.5-4.5 kpc and 3-6 kpc in G2E1, 0-3 kpc in G1E1, and 3-6 kpc in G1E0.5, the distributions of $n_2$ show peaks in their lower limits imposed by our prior. 

\begin{figure}[!ht]
\label{fig:parameter_profile}
\includegraphics[width=0.95\linewidth]{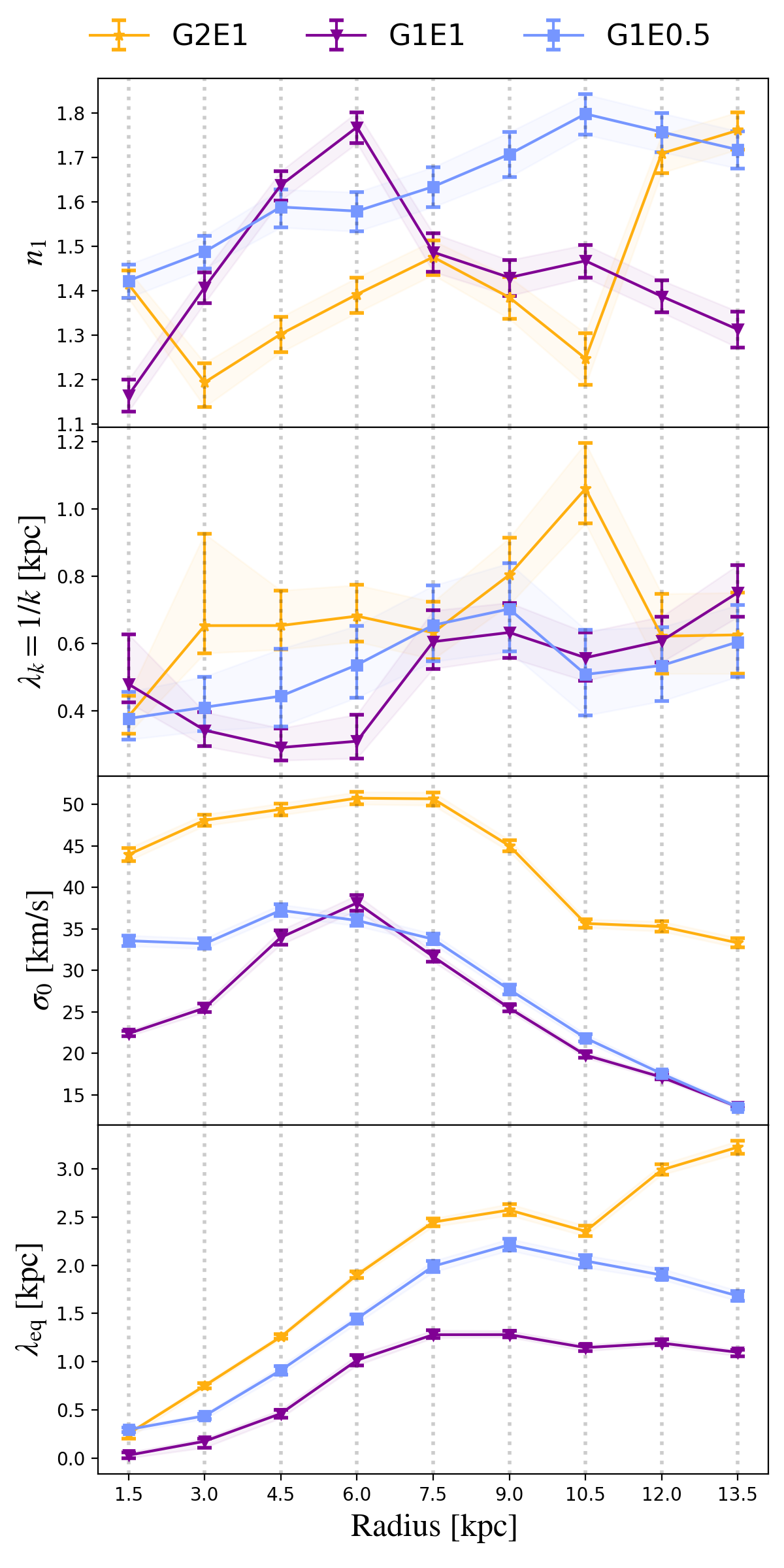}
\caption{Radial profiles of the median values of the model parameters $n_1$, $\lk=1/\kc$, $\sigma_0$ and $\lequi$. Each run is shown with a different color. Error bars represent the 16th - 84th percentile interval for each parameter.}
\end{figure}

Radial variations of the parameters are summarized in Figure \ref{fig:parameter_profile}. If we look at the first and second panels of Figure \ref{fig:parameter_profile} we can see that $n_1$ and $\lk$ are anticorrelated for runs G2E1 and G1E1. This is also true for G1E0.5 but is not noticeable in Figure \ref{fig:parameter_profile}. G1E1 and G1E0.5 show similar values of $\sigma_0$ at large galactocentric radius besides having different rotation curves. Their profiles of $\sigma_0$ have large magnitudes compared to the velocity dispersion profiles measured in nearby galaxies from CO emission lines \citep{Sun_2018}. For velocity dispersions derived from H I in \cite{Mogotsi_2016} our profiles are also higher. However, derived velocity dispersion profiles in \cite{Romeo_17} for some nearby galaxies show similar magnitudes, reaching up to 50 km s$^{-1}$ in H I and CO. In our model, $\sigma_0$ models the velocity dispersion of the whole annular region, with velocities measured at the maximum resolution and without a density cut. Then, $\sigma_0$ has not to be understood as the average velocity dispersion for clouds in an annular region.

\begin{figure*}
\label{fig:distributions_sims}
\includegraphics[width=\textwidth]{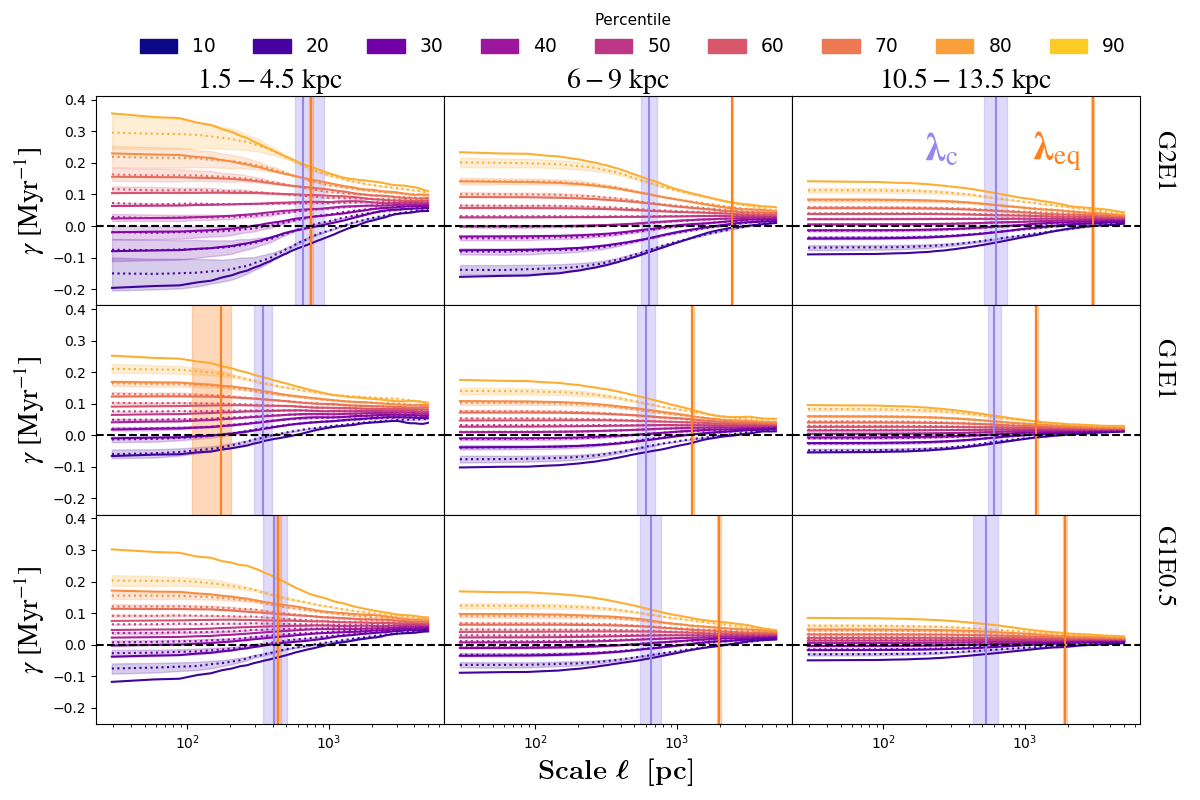}
\caption{Percentiles of circulation $\gamma$ at different annuli, as a function of scale $\ell$. Solid lines represent the percentiles of $\gamma$ in the simulation. Each color corresponds to a different percentile. Dotted lines show the percentiles of $\grot + \gno$ for the median values of  $n_1$, $\lk=1/\kc$, and $\sigma_0$. Shaded regions represent 1$\sigma$ uncertainty intervals from the posterior distributions of the model parameters. Vertical purple and orange lines show the spatial scales $\lk=1/\kc$ and $\lequi$, respectively, with their corresponding uncertainty illustrated by the shaded regions. Black dashed horizontal lines represent $\gamma=0$.}
\end{figure*}

\begin{figure*}
\label{fig:three-scales}
\includegraphics[width=\linewidth]{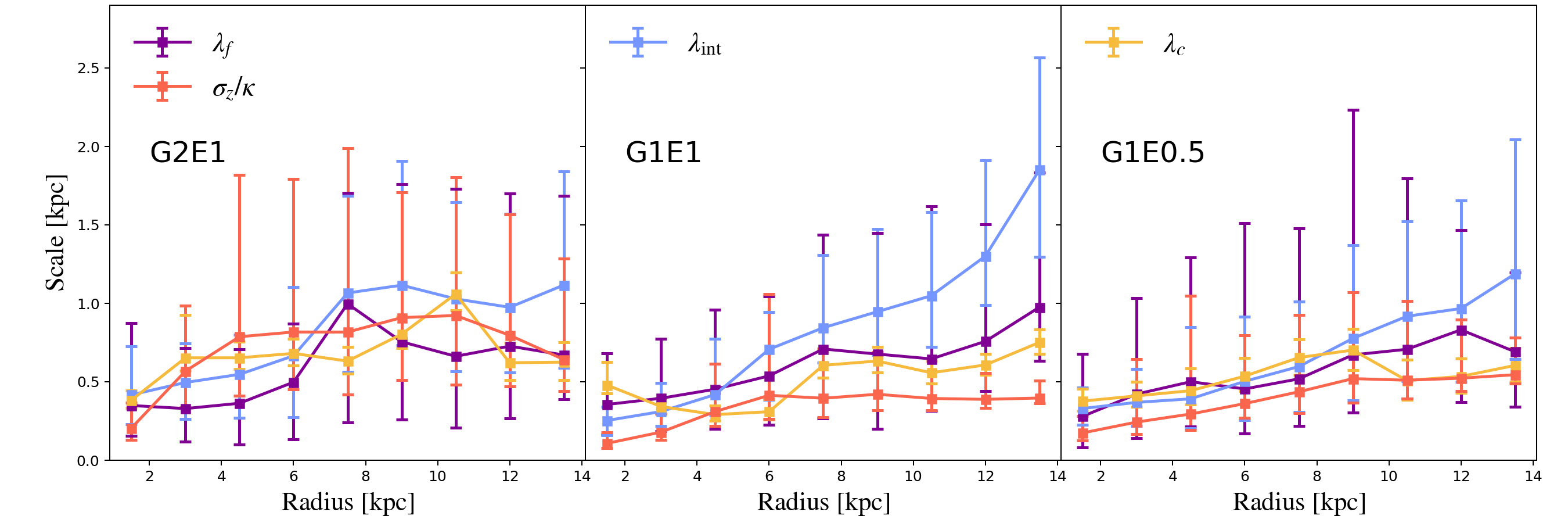}
\caption{Comparison between $\lk$ and characteristic spatial scales of gas clumps measured in the simulations. The orange line shows the epicyclic scale as a function of galactocentric radius. The purple line represents the length scale of fragmentation $\lambda_f =(M_c/\Sigma _{\rm gas})^{1/2}$. The blue line shows the characteristic separation between clumps, $\lambda_{\rm int}$. Error bars represent the 16th - 84th percentile interval for each parameter.}
\end{figure*}

\subsection{Scales at Which Gas Dynamics Transitions from Galactic Rotation to Noncircular Motions}

What is the role of galactic rotation on small scales? We want to know down to which scales galactic rotation still dominates the dynamics of gas or even molecular clouds. In our framework this information is encapsulated in the scale $\lequi$, the scale at which the contributions to the circulation field from galactic rotation and noncircular motions are roughly the same.
 
The bottom panel of Figure \ref{fig:parameter_profile} shows the radial profiles of $\lequi$. Within 8 kpc from the galactic center, $\lequi$ increases with galactocentric radius and varies between the resolution limit of 30 pc and 3 kpc. This shows that as we get farther from the galactic center, gas dynamics at the scale of clouds is predominantly dominated by noncircular motions.

In the radial annulus centered at 7.5 kpc, G2E1 and G1E1 have similar values of $\orot$, $n_1$, and $\lk$, while G2E1 has a higher $\lequi$ by about a factor of 2. This suggests that for the same rotation curve differences in $\lequi$ are mainly driven by differences in $\sigma_0$. The fundamental change between these two simulations is their gas surface density. Run G2E1 has the largest values of $\sigma_0$, and likewise it has the largest values of $\lequi$. On the other hand, G1E1 and G1E0.5 show similar profiles for $\sigma_0$ but $\lequi$ is larger for G1E0.5. This illustrates the effect of the rotation curve, which for G1E0.5 has a lower magnitude. In the central region of G1E1, $\lequi$ goes to zero, below the resolution of our simulations. This means that $\lequi$ is not resolved in these regions and that galactic rotation is the dominant source of circulation down to the resolution limit.

\subsection{Distribution of $\gamma$}

Now we show how the measured distributions of $\gamma$ change with spatial scale $\ell$ and how our model compares with them. Figure \ref{fig:distributions_sims} shows percentiles of the circulation $\gamma$ and the model $\grot +\gno$ as a function of scale for three of the nine annuli. Each percentile is shown with a different color. The percentiles of the measured circulation are shown as solid lines, while the models $\grot +\gno$ are shown by the dotted lines with their respective 1$\sigma$ intervals. 

Let us first discuss the general characteristics of these distributions. At the smallest spatial scale $\gamma$ is equal to the vorticity measured at the spatial resolution of the simulation. Near the galactic center the distributions of $\gamma$ are much broader at every scale $\ell$. This is also true for the large-scale component of circulation, $\grot$. As shown in Figure \ref{fig:rotation_curve}, the slope of $\orot$ decreases with radius, which means that within an annulus variations in $\orot$ also decrease with radius. We can see how the width of the distribution of $\gamma$ changes from large to small scales: toward small scales it gets broader as the influence of noncircular motions becomes more important, while above scales of hundreds of parsecs it starts to converge toward a constant level, set by the galactic rotation component. Figure \ref{fig:circulation_model_examples} in Appendix \ref{app:examples} shows how the distribution width of $\gamma$ looks for coherent rotation and a random field as a function of scale. 

Figure \ref{fig:rotation_curve} shows that $\omega_{\rm rot}$ and consequently the pdf of $\grot$ are always positive\footnote{according to the chosen orientation of the z-axis}. At galactic scales, $\gamma$ is greater than zero since $\gamma \simeq \grot$. On the other hand, the distribution of $\gno$, the GRF component, is half positive/half negative at any scale. At the scales where noncircular motions start to become important, $\gamma$ starts to show negative values. This departure to negative values is not the same for every region; it depends on the magnitude of $\grot$ and the dispersion of $\gamma$ at the smallest scales, which depends on $\gno$. By looking at the percentile curves, we see that the percentage of regions with retrograde rotation varies between 20\% and 40\% at the smallest scales, with the highest fractions in G2E1. 

We show the scales $\lk$ and $\lequi$  in Figure \ref{fig:distributions_sims}. For scales smaller than $\lk$ the rate at which the distribution broadens starts to decrease until it stops. This is also illustrated in the examples of GRFs in Figure \ref{fig:circulation_random_fields} in Appendix \ref{app:examples}. With regard to $\lequi$, we can see how $\lequi$ shifts from left to right depending on the average value of $\gamma$ at large-scales and its variance at the largest and smallest scales.

We see that our model reproduces the shape of the distributions of $\gamma$ as a function of $\ell$, with some discrepancies at both extremes of the distributions. Figure \ref{fig:distributions_sims} also displays $1\sigma$ uncertainties around the median value for our model of $\mV(k)$. Best agreement is seen at large galactocentric radii, where the distributions are better sampled since the number of cells in each annulus increases with radius. Near the galactic center we expect to observe large variations due to low sampling. In addition, it is more difficult to set a well-defined center of large-scale rotation near the center of the galaxy, since the interactions with small structures can be comparable to or greater than the large-scale gravitational influence of the galaxy. Regardless, Figure \ref{fig:distributions_sims} shows that our model can reproduce the trends in the measured distribution of $\gamma (\ell)$. This supports our assumption that the velocity field can be separated as the sum of large-scale circular motions and a GRF representing noncircular motions.

Although we are able to capture the general behavior of $\gamma$, there are noticeable discrepancies. In Figure \ref{fig:distributions_sims} we see a systematic discrepancy at the 90th percentile, with measured values greater than the model.  In some regions we also see discrepancies at the 10th percentile. In the first panel of G1E1 we see a less symmetry with respect to the median and a distribution that is broader than our model. This shows that the extreme values of $\gamma$ coming from noncircular motions are not represented in our model. One possibility is that within an annulus the model parameters change quickly. In our model, we are assuming a unique Gaussian distribution for each radial bin instead of a superposition of Gaussian distributions for each radius. However, in this scenario the distributions of $\gamma$ should be more symmetric. In some regions the deviation from the model of the 90th percentile is higher compared to the 10th percentile.

Some of the discrepancies could be explained by regions under collapse: when high-density regions collapse, their vorticity $\omega$ increases in magnitude.  \citet{Kruijssen_19} found that simulated molecular clouds with higher densities show higher velocity gradients. By checking the velocity divergence in the $x$-$y$ plane, i.e. $\nabla_{xy} \cdot \bvel = \dfrac{d v_x}{dx} +\dfrac{d v_y}{dy}$, we find that between the 20th and 80th percentiles of $\omega$ the median value of $\nabla_{xy} \cdot \bvel $ is close to zero. However, at the extremes of the distribution of $\omega$, $\nabla_{xy} \cdot \bvel $ drifts to negative values. Once a cloud of gas starts to collapse, $\nabla \cdot \bvel <0$, the magnitude of $\omega$ increases. It might be possible to address this discrepancy by considering the conservation of angular momentum or the conservation of circulation, but that is beyond the scope of this work.

\bigskip
\section{Discussion}
\label{sec:discussion}

We begin the discussion by commenting on the posterior distributions of the parameters $n_1$, $n_2$, $k_c$ and $\sigma_0$. Then, we discuss and interpret the scale $\lequi$. Before going into details, we have to remind that the properties of the velocity field derived from our parameters correspond to properties of the solenoidal component of the velocity field since $\omega$ and consequently $\gamma$ have no information of the irrotational component of the velocity field.

\subsection{Model Parameters}

We start by discussing the behavior of the exponents $n_1$ and $n_2$. The exponent $n_1$ shows how the circulation is distributed on larges scales down to the scale $\lk = 1/k_c$. Beyond $k_c$, i.e. at scales smaller than $\lk$, the function $\mV (k)$ quickly drops, showing values of $n_2$ with no apparent upper limits. As we show in Appendix \ref{app:power}, we get a similar result if $\mV (k) =0$ for scales larger than $\lk$. We also find in Appendix \ref{app:power}, that this break is necessary to reproduce the distribution of $\gamma$ at smaller scales. 

We have to point out that spatial correlations in the noncircular field are given by $\mV(k)$ and GRFs have coherent substructures unlike white-noise fields. The scale $\lk$ could be showing the size of the coherent structures in the velocity field \citep{Musacchio_17}. Coherent structures are long-lived structures that can be identified in the vorticity field \citep{Ruppert_05}, which transport mass and energy across different scales. For scales smaller than the size of these structures, i.e. $k>k_c$, the function $\mV(k)$ decays quickly, meaning that there is little information about the circulation field. One interpretation is that circulation or rotation is transferred from large scales down to the scale $\lk$. Below the scale $\lk$ the redistribution of circulation from large scales stops and the distribution of $\gamma$ starts to converge. At these scales the solenoidal component of the velocity field starts to show coherent structures that are decoupled from the random behavior of the noncircular component. An observational example of this scale might be found in \cite{Rosolowsky_2003}, where the velocity gradients of massive of clouds within regions 500 pc are preferentially aligned. \cite{Rosolowsky_2003} show an observational correlation in the velocity field for scales smaller than 500 pc, which is similar to the values we find for $\lk$.

The next step is to link the structures in the velocity field with structures in the density field. One of the first scales that should affect the behavior of the velocity field is the scale height. However, the scale height of the gas density field for the three runs is of the order of 100 pc, which is usually smaller than $\lk$ by a factor of 4-5. We continue analyzing different structures that exist in the plane of galactic disks.

Let us assume that the scale $\lk$ is related to the fragmentation of gas with a characteristic scale $\lambda_f$. If $\Sigma_{\rm gas}(R)$ is the gas surface density profile, we can estimate a characteristic mass for the fragments or clumps as $M_c \approx \pi \Sigma_{\rm gas} (\lambda_f/2)^2$, and then the scale of fragmentation is approximated by
\begin{equation}
\lambda_f \approx \left(\frac{M_c}{\Sigma_{\rm gas}}\right)^{1/2}.    
\end{equation}

The scale $\lk$ could also be related to interactions between clumps \citep{Dobbs_2013}; however, this interpretation is not straightforward since we do not find an energy cascade at smaller scales. If cloud-cloud interactions play a role, it is worth studying the typical distance between clumps of gas at each radius.\\

To compute the two aforementioned scales, we need to define a clump of gas. We use the following criteria to define a gas clump:
\begin{itemize}
    \item Each cell in the clump has a number density above $100$ cm$^{-3}$.
    \item Clumps are gravitationally bound.
    \item Clumps have a minimum of 20 cells (a $3\times 3 \times 3$ cube without the corners has 19 cells).
\end{itemize}

The clumps identified by these criteria have typical sizes of 100 pc, while in nearby galaxies sizes range between 10 and 100 pc \citep{Rosolowsky_07,Heyer_09,Colombo_14}. These sizes are of the order of the scale height across the disk for the three runs. We measure $M_c$ as the total gas content in the clump. To compute the typical distance, we do the following: for each clump we average the distance to the three closest clumps, and we average that quantity at each annulus. We refer to this scale as $\lambda_{\rm int}$\\ 

We show $\lambda_f$ and $\lambda_{\rm int}$ in Figure \ref{fig:three-scales}. For most regions, $\lambda_{\rm int}$ and $\lambda_f$ are of the same order of magnitude, and they appear to be correlated. The spatial scale $\lk$ seems to lie closer to $\lambda_f$. However, the scatter of these scales is too large to derive any strong conclusion. Each annulus has a characteristic $\Sigma_{\rm gas}$, $\Omega$, and $\sigma _v$, quantities that are dynamically correlated. In that regard it is not surprising that spatial scales defined by these quantities show similar behaviors. 

Another scale related to a change of the behavior of the velocity field is the epicyclic scale $\sigma_v/\kappa$ \citep{Meidt_2018}, where $\sigma_v$ is the velocity dispersion of a gas cloud and $\kappa$ is the epicyclic frequency. Epicyclic motions correspond to the evolution to small perturbations of circular orbits under the gravitational potential of the galaxy. Structures larger than their corresponding epicyclic scale are ensured to be affected by the galactic potential. Figure \ref{fig:three-scales} shows $\sigma_z/\kappa$ as the orange solid line, where $\sigma_z$ corresponds to the dispersion velocity in the $z$-axis in a radial bin. The choice of $\sigma_v=\sigma_z$ assumes that once a structure has formed, its velocity dispersion is nearly isotropic. Like the scales $\lambda_f$ and $\lambda_{\rm int}$, the epicyclic scale lies close to $\lk$. 

The physical correlations between all these scales and their level of uncertainty make it difficult to compare them with $\lk$. Therefore, we are not able to elucidate the fundamental physical origin of $\lk$. We can only conclude that $\lk$ is related with the formation of structure in our simulations and that the details of gas dynamics below such structures do not significantly affect the overall circulation of gas.

\subsection{Circulation Scales: $\lk$ and $\lequi$}

\begin{figure*}
\label{fig:equi}
\includegraphics[width=\linewidth]{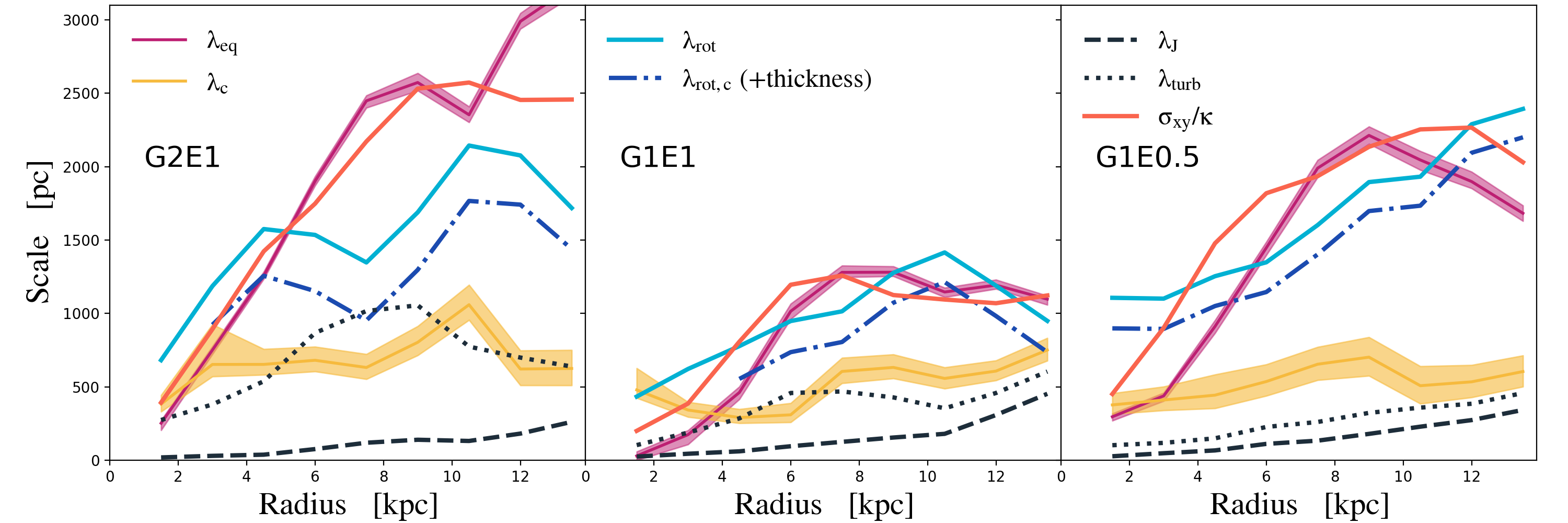}
\caption{Spatial scales as a function of galactocentric radius: solid pink and yellow lines correspond to $\lequi$ and $\lk$ respectively. The shaded regions correspond to 1$\sigma$ uncertainties. The classical instability scales for two-dimensional disks, $\lrot$ and $\lj$, are shown as solid light-blue lines and black dashed lines, respectively. The dotted-dashed line shows the effective values $\lrot$ after adding the effects of resolution.}
\end{figure*}

We start this Section by discussing the role that gravitational instabilities can play in setting the distribution of circulation, particularly their effect in $\lk$ and $\lequi$. First, we recall the Toomre parameter, $Q =\kappa \sigma_v/\pi G \Sigma$, which for marginal stable systems ($Q \approx 1$) and with constant $\kappa$ requires $\sigma_v \propto \Sigma$. At the scales of clouds, for virial parameters $\alpha \approx 2$, $\sigma_v$ also grows with $\Sigma$ ($\sigma_v \propto \sqrt{\Sigma_{\rm gas}}$) \citep{Sun_2018}. For any of these two pictures, we expect that galaxies with more gas have more randomness in their velocity fields. This is illustrated by the run G2E1 in Figure \ref{fig:parameter_profile}, which shows higher values of $\sigma_0$ and $\lequi$ compared to the other runs. 

Since we are dealing with galactic disks, the first step to visualize relations with gravitational instabilities is to compute $\lrot$, and the two-dimensional thermal Jeans scale $\lj$, given by $c_s^2/G\Sigma $, where $c_s$ is the gas sound speed and $\Sigma$ is the gas surface density. The thermal Jeans scale sets the size of the smallest structures that can be formed. Both length scales, $\lrot$ and $\lj$, are shown in Figure \ref{fig:equi}.\\

Since the two-dimensional stability is affected by the disk thickness (or the resolution in the case of simulations), we have to consider the dispersion relation $\omega_p ^2(k) = \kappa^2 -2\pi G \Sigma \vert k \vert e^{-k \epsilon}$, where $\omega_p$ is the frequency of perturbations and $H$ is the disk thickness with a minimum value set by the numerical resolution (\citealp{Binney_Tremaine_08}, p. 552). Perturbations where $\omega_p ^2(k) <0$ correspond to instabilities. To obtain the correct values if $\lrot$, which we call $\lrotc$, we solve the equation $\omega_p ^2(k)=0$. We plot $\lrotc$ in Figure \ref{fig:equi} as the dotted-dashed blue line. Once the disk thickness is considered, near the galactic centers $\omega_p$ is always real and any radial perturbation is stable. This means that in the inner regions of these galaxies gravitational instabilities are not resolved. At these scales we do not expect to see an important injection of energy due to gravitational instabilities. The effect of the disk thickness is more noticeable in our additional simulations in Figure \ref{fig:equi_app}: if $\lrotc$ is not resolved, $\lequi$ falls below the resolution of the simulations.

For this set of simulations, $\lk $ lies between $\lj$ and $\lrotc$, that is, within the range of scales in which gravitational instabilities can exist, consistent with this scale being associated with scales of structure formation. For all runs we see that $\lequi$ is above $\lj$, it increases with radius, and in some regions is higher than $\lrotc$. Since $\lrotc$ is the maximum size of unstable perturbations, clouds formed in regions where $\lequi > \lrotc$ will be predominantly dominated by noncircular motions.

For all runs we see that $\lequi$ can show values up to kiloparsec scales. For G2E1, $\lequi$ is higher than $\lrot$ in most regions. This is probably caused by the high star formation rate, due to its higher gas content and associated increase in feedback-induced noncircular motions. In Figure \ref{fig:equi_app} in Appendix \ref{app:simulations}, our simulations with only SN feedback show values of $\lequi$ lower than $\lrot$ and lower than $\lk$. This last point shows that only considering the difference in gas content between simulations without taking into account stellar feedback is insufficient to explain differences in $\lequi$ across the different runs. These additional simulations also show an apparent correlation between $\lequi$ and $\lrot$. However, once we use a more energetic type of feedback $\lequi$ grows, and this apparent correlation disappears. This suggests that there might be two different regimes where the distribution of circulation is set by gravitational instabilities or by stellar feedback. This idea goes in line with results from numerical simulations and analytical models showing that turbulence can be powered by gravity or stellar feedback, and that the dominant driver of turbulence changes across the evolution of the universe \citep{Krumholz_10,Goldbaum_15,Goldbaum_2016,Krumholz_2018}.

Since we are discussing the turbulent behavior of gas and its dynamical stability, we can also discuss the relevance of the turbulent Jeans scale $\lambda_{\rm turb} = \sigma_v^2/G\Sigma$, where $\sigma_v$ is the velocity dispersion of gas considering nonthermal motions. However, $\sigma_v$ is a function of scale $\ell$ \citep{Elmegreen_2004,Romeo_2010}, and to properly take into account the effects of turbulence, we need to know how $\sigma_v$ changes with $\ell$. Here we take a first-order approach considering $\sigma_v = (c_s^2 + \sigma_z^2)^{\frac{1}{2}}$, where $\sigma_z$ is the mass-weighted vertical dispersion velocity in the disk of the galaxy. In this approximation we are assuming that the velocity dispersion at the scale of the disk sale height is a representative value of the velocity dispersion for bound structures in the presence of turbulence. To compute $\lambda_{\rm turb} $ we use a temperature cut of $T<$ 5000 K to avoid considering gas that is currently affected by stellar feedback. We show $\lambda_{\rm turb} $ in Figure \ref{fig:equi}. For runs G2E1 and G1E1 $\lambda_{\rm turb}$ is of the order of $\lk$ which suggests that $\lk$ could be tracing the scales at which turbulent structures are affected by their self-gravity. This is not shown by run G1E0.5, but we need to keep in mind that we are assuming that $ \sigma_z^2$ is a good proxy for the turbulent velocity in the plane of the galaxy for self-gravitating structures.

In the field of fluid dynamics, it is known that in turbulent fluids coherent structures naturally arise, and that these structures are fundamental for the transport of angular momentum across different scales \citep{Kraichnan_67,Ruppert_05}. This motivates us to look for structures that can be defined by kinematics only. One alternative is to look for structures whose behavior is defined by the ratio between the galactic angular velocity, which traces the galactic potential, and the local noncircular motions at cloud scales. A spatial scale that goes in that direction and compares the magnitudes of the velocity dispersion of gas and the galactic potential is the epicyclic scale $\sigma_v/\kappa$ \citep{Meidt_2018}. However, as shown in Figure \ref{fig:three-scales}, if we use the velocity dispersion across the $z$-axis, the epicyclic scale is of the order of $\lk$, which in most regions is smaller than $\lequi$. Another approach is to consider the velocity dispersion in the $x-y$ plane, $\sigma_{xy}$, within a radial bin. To compute $\sigma_{xy}$ we subtract the circular velocity model from the velocity field. The scale $\sigma_{xy}/\kappa$ compares the energy in the noncircular velocity field with respect to epicyclic motions given by the galactic potential. We show the spatial scale $\sigma_{xy}/\kappa$ in Figure \ref{fig:equi} as the orange line.   

Figure \ref{fig:equi} shows that $\sigma_{xy}/\kappa$ is similar to $\lequi$. This result would suggest that the ratio $\sigma_{xy}/\kappa$ is a good proxy for $\lequi$. However, this might be valid only for our feedback prescription and for galaxies with an average star formation rate according to the Kennicutt-Schmidt relation \citep{Kennicutt_98, Daddi_10}. For our second set of simulations described in the Appendix \ref{app:simulations} with only SN feedback, the values of $\lequi$ are usually lower than $\sigma_{xy}/\kappa$ and $\lk$. This again shows the effect of using different feedback prescriptions. Momentum feedback prescriptions change the velocity field more aggressively; $\lequi$ shows large values and is similar to $\sigma_{xy}/\kappa$. 

We can interpret this difference as differences in the sources of noncircular motions. Simulations with only thermal SN feedback can lose this source of energy quickly due to our resolution of 30 pc, producing a lower effect in the velocity field. In this scenario, gravitational instabilities might become a relevant source of noncircular motions or turbulence. On the other hand, the stellar feedback prescription used in our main simulations changes explicitly the velocity field at the smallest scales and increases directly the magnitude of $\mV (k)$. Feedback might erase the correlation between gravitational instabilities and the noncircular motions at small scales.

This point has implications for the analysis of star formation in numerical simulations. In simulations with mechanical stellar feedback, i.e. momentum injection, the velocity field is explicitly changed and the amount of kinetic energy at small scales increases. This reduces the coupling between the dynamics of clouds and galactic rotation, as well as the coupling between the efficiency of star formation and the galactic environment. The magnitude of correlations between star formation and galactic properties found in simulations might depend on the specific stellar feedback prescription used.

Up to this point we have shown that the distribution of circulation is affected by stellar feedback and gravitational instabilities. It is interesting to discuss what other studies show with respect to the distribution of circulation or rotation. Here we mention the works of \cite{Tasker_09} and \cite{Ward_2016}, which use simulations to measure how the rotation of molecular clouds aligns with respect to the rotation of their galaxies. These simulations have similar surface gas densities and the same shape of the velocity curve, but with different magnitudes. Both simulations have weak forms of feedback; \cite{Tasker_09} did not include stellar feedback in their simulations, and the simulations in \cite{Ward_2016} had a reduced feedback efficiency of 10\%. Hence, we expect that their distribution of circulation is set by gravitational instabilities. For comparison, at a galactocentric radius of 8 kpc, $\lambda_{\rm rot}\simeq 2100$ pc in \cite{Tasker_09} and $\lambda_{\rm rot}\simeq 1500$ pc in \cite{Ward_2016}. In addition, the simulation of \cite{Ward_2016} shows spiral structures, while the density field in \cite{Tasker_09} is more random. The simulation of \cite{Tasker_09} is more unstable than the one from \cite{Ward_2016}, and consequently the former should have higher values of $\lequi$, or a higher fraction of molecular clouds with retrograde rotation with respect to their galaxy. These simulations effectively find different fractions of retrograde clouds, 30\% in \cite{Tasker_09} and 13\% in \cite{Ward_2016}. This shows that more unstable systems are more dominated by noncircular motions and have higher values of $\lequi$. \cite{Tasker_09} also analyzed the effects of resolution, which directly influences the size of molecular clouds and the stability of gas dynamics as shown by $\lrotc$. \cite{Tasker_09} show that as the resolution is increased, more molecular clouds present retrograde rotation. In summary, these studies show that in the absence of strong feedback, gravitational instabilities play a role in setting how circulation is distributed at smaller scales and the relevance of the spatial resolution used in numerical simulations.

\subsection{Turbulence and the Power Spectrum}

Since we are studying two-dimensional velocity fields we discuss how our results are compared with known properties of two-dimensional turbulence. A turbulent velocity field $\delta \bvel$ is characterized by its kinetic energy spectrum $E(k)$ such that the mean turbulent kinetic energy per unit mass is $\frac{1}{2}\langle \delta v^2 \rangle = \int_0 ^{\infty} E(k) dk$. The power spectrum $E(k)$ of two-dimensional turbulence is characterized by the existence of two inertial regimes: (i) an inverse energy cascade $E(k) \propto k^{-5/3}$ for $k<k_f$ and (ii) a direct enstrophy cascade with $E(k) \propto k^{-3}$ for $k>k_f$, where $k_f$ is the wavenumber of the forcing scale \citep{Kraichnan_67,Wada_02,Bournaud_10,Musacchio_17}. In our model, the random component of the velocity $\vno$ field is characterized by $\mV(k)$. From the Appendix \ref{sec:random_fields} it can be shown that $E(k)\propto \mV(k)^2 k$ for a two-dimensional field, which is the case of $\vno$.
For two- and three-dimensional turbulence, the relation between $n$ and the energy spectrum $E(k)$ is given by
\begin{equation}
\text{(2-D)} \qquad  E(k) \propto v(k)^2 k \propto k^{-2n+1}. 
\end{equation}

In this formalism, the inverse energy and the direct enstrophy cascade are represented by $n=4/3 \approx 1.33$ and $n=2$. As shown in Figure \ref{fig:parameter_profile}, the distribution for the exponent $n_1$ ranges between 1.1 and 1.8. This exponent lies close to the expected values of both regimes. On the other hand, $n_2$, which should be associated with the enstrophy cascade, has no upper limit in our model, and the scale where $\mV(k)$ breaks, $\lk$, is of the order of 240 pc to 1 kpc. If we look at the middle panel of Figure \ref{fig:circulation_random_fields} in Appendix \ref{app:examples}, we see that for high values of $n$ the distribution of $\gamma$ starts to be less sensitive to changes in $n$. It is likely that most of the information of the distribution of $\gamma$ is given by $n_1$ and $k_c$. If this is the case, $k_c$ is related to the turbulent forcing scale $k_f$. Experiments of thin layer fluids show strong long-lived vortices at the scale $\lambda_f=1/k_f$ \citep{Musacchio_17}. Then, the turbulent picture also suggests that $\lk$ might be related to the formation of structure in the turbulent velocity field.

Numerical and observational studies suggest that information about the turbulent velocity field can be extracted from the spectrum of the gas surface density \citep{Elmegreen_01,Combes_2012,Bournaud_10}. The numerical work of \cite{Bournaud_10} shows that the spectrum of the surface density field $\Sigma (k)$ may be described by a broken power law with a critical scale that is interpreted as the disk scale height. The slope of the power law in \cite{Bournaud_10} simulations changes from -2 at large scales to -3 small scales. Measuring the density spectrum $\Sigma (k)$ for our simulations, we find that the slope changes from around zero for scales larger than 500 pc (about the same order of $\lk$) to a continuous decaying function at small scales with no clear critical scale. In our simulations we are not able to link the density and the power spectrum of gas.

\section{Limitations and Caveats}
\label{sec:Limitations}
The work presented here is largely exploratory and aimed at establishing the basic concepts associated with modeling the spatially dependent distribution of gas circulation in disk galaxies. Here we discuss some of the limitations associated with this modeling. In a future work we expect to address several of these limitations in order to apply the presented methods to extract information from observed galaxy velocity fields.

\subsection{Velocity Model}
The main assumption in our model is that the velocity field in the plane of the disk can be approximated as the contributions of two different fields: $\pmb{V}=\pmb{V}_{\rm rot}+\pmb{V}_{\rm nc}$, where $\pmb{V}_{\rm rot}$ corresponds to the galactic velocity curve and $\pmb{V}_{\rm nc}$ corresponds to a Gaussian random velocity field. In real galaxies, we find other types of coherent motions that are different from galactic rotation and pure random motions. Among these, we find induced motions by galactic bars and spirals, and epicyclic motions. At the scale of epicycles, gas is still affected by the tidal forces exerted by the galactic potential \citep{Meidt_2018}. According to \cite{Meidt_2018}, depending on the strength of self-gravity, the dynamical structure of clouds shows preferred orientations in radial or azimuthal coordinates, which does not occur in our model of $V_{\rm nc}$. In addition, galactic bars and spirals would also produce deviations from global galactic rotation, which we are implicitly including in $\pmb{V}_{\rm nc}$.

Figure \ref{fig:distributions_sims} shows that our simple model can successfully fit the distribution of circulation across different spatial scales in general terms, but there are some clear deviations in particular regimes (e.g., small-scale, prograde rotating regions with high values of $\gamma$ at intermediate galactocentric radii), which probably signal more complex types of motions not recovered by the model. Moreover, due to conservation of angular momentum, the vorticity in high-density regions is enhanced. It is unclear how to statistically model these types of motions.

\subsection{Full Velocity Field}
In this work we have made use of isolated galaxy simulations whose rotation axis is aligned with the z-axis of the simulation box by default. To compute the circulation, we used the two-dimensional velocity field $\pmb{V} (x,y)$, the density-weighted projection across the z-axis of the three-dimensional velocity field. We can separate the total circulation into two terms, $\Gamma_x$ and $\Gamma_y$:

\begin{equation}
\displaystyle \Gamma = \oint \bvel \cdot d\pmb{r} = \oint v_x \cdot dx + v_y \cdot dy = \Gamma_x + \Gamma_y,
\end{equation}

where $x$ and $y$ are coordinates on the plane of the disk.
In observations we only have access to the velocity along the line of sight, which will be the sum of one of the velocities in the plane of the galaxy, $v_x$ or $v_y$ and $v_z$, motions vertical to the disk midplane.
Consider a disk with inclination $i$ such that the line of sight lies in the $x$-$z$ plane. The coordinates in the plane of the sky are $z'=z\sin i +x \cos i$ and $y'=y$. The axis of the line of sight is $x'=x\sin i + z\cos i$, and the velocity is $v_{\rm LOS}=v_x'=v_x \sin i + v_z \cos i$. On the midplane $z=0$  and the projected position $z'=x\cos i$. From the observed quantities we can compute 
\begin{equation}
\label{eq:observed}
\begin{split}
    \Gamma' &= \oint v_{\rm LOS} dz' = \sin i \cos i \oint v_x dx +  \cos^2 i \oint v_z dx\\
    \Gamma' &= \sin i\cos i \Gamma_x + \cos^2 i K_{xz},
\end{split}
\end{equation}
where $\Gamma_x$ is the component of $\Gamma$ in the $x$-axis and $K_{xz}$ is the sum of vertical motions along the $x$-axis. The term $K_{xz}$ should be of the order of $\langle v_z\rangle \ell$. This component has to be treated as an additional term in the assumed decomposition of the velocity field. 
In this work we have assumed that the velocity field in the plane is the sum of galactic rotation plus a random component. For $v_x$ this means $v_x=-\Omega y + \delta v_x$, where $\Omega$ is the galactic angular velocity and $\delta v_x $ is the random velocity term. For $v_z$ we can assume that $v_z=\delta v_z$. Then, $ \Gamma' \approx  -\cos i[y \sin i \oint \Omega dx + \ell (\langle v_x \rangle \sin i- \langle v_z \rangle \cos i) ]  $
At large scales $\langle v_x \rangle$ and $\langle v_z \rangle$ are approximately zero, while at small scales both terms behave like random variables. 
The sum of these two terms would be the observed random component. Since we want to compare them with galactic rotation, the best inclination has to maximize the contribution of $\Omega$ to $\Gamma'$ that occurs at $i=45^\circ$.

\subsection{Surface Brightness Limits and Recovery of Velocity Information}

A major limitation comes from the observational detection limits for different transition lines, which lead to an incomplete sampling of the velocity field. For example, the CO (1-0) transition has a critical density $\simeq 10^3 \rm cm^{-3}$, tracing the distribution of molecular gas in galaxies. This implies that we can only observe a small fraction of the velocity field at the scales of molecular clouds. Proper ways of dealing with noise and censored data in faint regions will also need to be implemented.

\subsection{Simulations}

In this paper we use hydrodynamical simulations of disk galaxies to test our method. The results presented here are valid to our set of simulations, with their defined prescriptions for star formation and stellar feedback. However, caution should be taken before directly extrapolating our findings to the environments of real galaxies.  Here we list what we consider are the most important aspects in which our simulations and observed galaxies differ:

\begin{itemize}
 \item Resolution: The maximum spatial resolution corresponds to 30 pc. In practice, this means that we are able to resolve structures and instabilities of the order of 100 pc, corresponding to approximately four times our resolution. In nearby galaxies, the size of molecular clouds typically ranges from tens to hundreds of parsecs. Although we see formation of structures, this resolution is not enough to resolve the inner turbulence of molecular clouds, their gravitational collapse, and the interactions of clouds smaller than 100 pc. A higher resolution would imply more interactions and a higher velocity dispersion at the smallest scales studied here. Then, we might expect a change in the values of $n_2$ that sets the behavior of $\mV(k)$ at large wavenumber $k$, i.e. at smaller spatial scales. Despite this caveat, $\lequi$ and $\lk$ are well resolved almost everywhere.
 \item Temperature: Gas is allowed to cool owing to radiation down to a temperature of 300 K. This means that the smallest structures in our simulations are more similar to H I clouds. Also, this temperature floor sets a minimum Jeans scale as a function of gas surface density 
 \begin{equation}
 \lambda_{J,{\rm min}}= \dfrac{5}{3} \dfrac{k_B T_{\rm min}}{\mu m_p}  \dfrac{1}{G\Sigma_{\rm gas}} = 133 {\rm pc}  \left(\dfrac{\Sigma_{\rm gas}}{10 \msun {\rm pc}^{-2}}\right)^{-1}
 \end{equation}
assuming a mean molecular weight $\mu=0.6$.
In local galaxies, the Jeans length is of the order of a few parsecs, about two orders of magnitude below the average Jeans scales found in our simulations. The temperature floor leads to an overestimation of the relevance of the Jeans scale. It is important to mention that to compute the radial profiles of $\lj$ shown here, we are considering all the gas in an annulus and its respective average temperature instead of the average $\lj$ for cold and dense gas. This makes sense for our analysis since we are computing the circulation for all the gas within the physical volume described in the paper. However, for regions with densities below $10 \msun {\rm pc}^{-2}$ the minimum value for $\lj$ is larger than four resolution elements. Then, even for our resolution the values of $\lj$ are likely overestimated and should be considered as upper limits.

\item Stellar feedback: In our recipe of stellar feedback, we include the direct injection of momentum from radiation pressure and stellar winds to the six nearest cells. Although the amount of added momentum $\pmb{p}_{\star}$ does not explicitly change with spatial resolution, the typical masses, $m_{\rm cell}$, of cells around star particles do change with different spatial resolution. This translates in different magnitudes for the change of the velocity field around star particles, since the velocity $\bvel_{\star}=\pmb{p}_{\star}/m_{\rm cell}$. We have not tested how sensitive to resolution is this feedback prescription.

\item Spiral arms and bars : A relevant difference between observations, other simulations, and our runs is that our simulated galaxies lack grand-design spiral arms. The main difference is that the old stellar population in this work is represented by an external axisymmetric potential, whereas other studies use particles \citep{Renaud_13} or spiral potentials \citep{Dobbs_15}. Only new stars are particles; hence, only this stellar component can respond to perturbations making the stellar disk more stable.
The impact of spiral arms and bars in our analysis can be separated by their effect on large and small scales. At large scales, the bulk motion $\vrot$ must be a function of radius and the azimuthal angle. At small scales, spiral and bars can induce vortex motions by Kelvin-Helmholtz instabilities, Rayleigh-Taylor instabilities, or tidal fields \citep{Dobbs_Bonnell,Renaud_13}. These structures create new sources of turbulence; therefore, the velocity field at the scale of molecular clouds has different properties. In this work, we have divided disk galaxies into annular regions and measured radial profiles of the parameters that define the small-scale velocity field $\vno$. To test the effect of spirals and arms, we also need to separate regions according to their azimuthal distance to these structures.

\end{itemize}

\section{Summary and Conclusions}
\label{sec:conclusions}
In this study we characterize the rotation of gas in galaxies at different scales by measuring the circulation $\Gamma$, a macroscopic measure of fluid rotation. We develop a method to measure the contributions of large-scale motions, i.e. galactic rotation, and noncircular motions in the observed distribution of circulation at different spatial scales. Noncircular motions are modeled as random Gaussian velocities described by a generating function in Fourier space $\mV(k)$.
We apply this method on three hydrodynamical simulations of galactic disks, performed with the AMR code Enzo, which includes star formation, SN feedback, and momentum feedback from stellar winds, and with a spatial resolution of 30 pc.

We summarize the major points of this work:

\begin{itemize}

    \item We model the velocity field of galaxies with two components: a galactic component given by the circular velocity profile, and a Gaussian random component. The random component is obtained from a function $\mV(k)$ whose functional form corresponds to a broken power law with exponents $n_1$ and $n_2$, transitioning at the wavenumber $k=k_c$. The amplitude of $\mV(k)$ is defined by the characteristic velocity dispersion of the random field $\sigma_0$. We apply the model to hydrodynamical simulations and confirm that motions can be well modeled by two components with different circulation, as hypothesized. The model successfully reproduces the distribution of circulation as a function of scale, except when regions are under gravitational collapse.
    
    \item We find that a sharp transition in the behavior of gas dynamics at the scale $\lk=1/k_c$ is necessary to fit the circulation distribution. This may correspond to the scale at which kinematics transition from being coupled to the galaxy to more disordered motion, associated with feedback-driven turbulence or gravity-driven turbulence. However, the resolution of the current simulations limits our ability to probe this in greater detail. 
    
    \item The scale $\lk$ is similar to the scale at which gas fragments and to the epicyclic scale $\sigma_z/\kappa$ that defines the scale at which self-gravity and the potential of the galaxy are equally important to determine the internal dynamics of clouds. The scale $\lk$ is also similar to the scale of fragmentation and the distance between clumps, suggesting that $\lk$ shows the formation of structure in the density field. 

    \item We introduce a dynamical spatial scale $\lequi$. At spatial scales similar to $\lequi$ the contributions of galactic circular motions and noncircular motions to the observed circulation or local rotation of gas are roughly the same. For regions larger than $\lequi$ galactic rotation dominates the circulation of gas and consequently the measured rotation. At these scales the distribution of circulation shows largely positive values, which means that gas rotates in the same orientation of the galaxy. For patches of gas smaller than $\lequi$, noncircular and random motions start to dominate the observed circulation and retrograde rotating regions can be found.
    
    \item We find that $\lequi$ depends on the local properties of gas. From the center of the galaxies, $\lequi$ increases with galactocentric radius. This shows that the spatial scale at which gas dynamics is dominated by noncircular motions depends on the position in the galactic disk. We see different behaviors in the central regions and outskirts of galaxies. Galactic rotation appears to be more important or dominant toward the center of galactic disks. 
    
    \item The scale $\lequi$ is similar to the ratio $\sigma_{xy}/\kappa$, as predicted by models about balance of rotation and turbulence, where $\sigma_{xy}$ is the in-plane velocity dispersion of the two-dimensional velocity field within a radial annulus and $\kappa$ is the epicyclic frequency. However, when suppressing momentum feedback in simulations, $\lequi$ can be lower than $\sigma_{xy}/\kappa$, and its radial profile might be correlated with the spatial scales of gravitational instabilities since self-gravity becomes a relevant source of turbulence.
    
    \item In some regions $\lequi$ is greater than $\lrot$. The formation of structures in such regions will be dominated by noncircular motions. Depending on the sources of feedback, turbulence, and local dynamics, $\lequi$ can be larger or smaller than $\lrot$. Particularly, for strong modes of stellar feedback $\lequi$ can be larger than $\lrot$.
    
    \item Stellar feedback changes $\lequi$ by injecting momentum at smaller scales, increasing motions that do not follow galactic rotation. Different prescriptions of stellar feedback will produce different velocity fields that show different coupling between the dynamics of gas at small scales and the large-scale galactic rotation. This can also produce changes in the coupling between star formation and galactic rotation.
    
\end{itemize}

This works shows that the characterization of the ISM circulation, from the modeling of velocity fields of galaxies, opens the possibility of directly measuring scales associated with gravitational collapse and structure formation and studying how it changes with galaxy properties and local conditions. It also shows that rotation is dynamically important in some environments like the centers of galactic disks. In the future, the analysis of circulation can be extended to real galaxies to study the relevance of galactic rotation at the scale of molecular clouds. 


\acknowledgments

Some of the computing for this project was performed on the Memex cluster. We would like to thank Carnegie Institution for Science and the Carnegie Sci-Comp Committee for providing computational resources and support that contributed to these research results. Powered@NLHPC: This research was partially supported by the supercomputing infrastructure of the NLHPC (ECM-02). The Geryon/Geryon2 cluster housed at the Centro de Astro-Ingenieria UC was used for part of the calculations performed in this paper. The BASAL PFB-06 CATA, Anillo ACT-86, FONDEQUIP AIC-57, and QUIMAL 130008 provided funding for several improvements to the Geryon2 cluster.
J.U. acknowledges support from Programa Nacional de Becas de Postgrado, CONICYT (grant D-21140839). A.E. acknowledges support from CONICYT project Basal AFB-170002 and Proyecto FONDECYT Regular grant 1181663. F.B. acknowledges funding from the European Research Council (ERC) under the European Union’s Horizon 2020 research and innovation program (grant agreement No. 726384). J.M.D.K. gratefully acknowledges funding from the German Research Foundation (DFG) in the form of an Emmy Noether Research Group (grant No. KR4801/1-1) and from the European Research Council (ERC) under the European Union's Horizon 2020 research and innovation programme via the ERC Starting Grant MUSTANG (grant agreement No. 714907). A.H. is a fellow of the International Max Planck Research School for Astronomy and Cosmic Physics at the University of Heidelberg (IMPRS-HD).

\appendix

\section{Vorticity and Angular Momentum}
\label{app:angular}
\label{sec:vorticityangularmomentum}
Let us approximate the velocity field around a point $\pmb{r}_0=x_0\hat{x}+y_0\hat{y}$ in Cartesian coordinates. We choose to do this analysis in Cartesian coordinates since cylindrical or spherical coordinates system are noninertial reference frames. The unit vector $\hat{x}$ is instantly aligned with the radial unit vector $\hat{R}$, i.e. $\hat{x}=\hat{R}$ and $\hat{y}=\hat{\phi}$, where $\phi$ is the azimuthal cylindrical coordinate. Then, $x_0=R_0$ and $y_0=0$. For a velocity field $\bvel=R\Omega\hat{\phi}$,
\begin{eqnarray}
v_x =& -y_0 \Omega(R_0)&=0\\
v_y =& x_0 \Omega(R_0)&=R_0\Omega(R_0). 
\end{eqnarray}
We expand $\Omega(R)$ in terms of $x$ and $y$, around the point ($R_0+x,0+y$). Since the Cartesian coordinates are aligned with the cylindrical coordinates $R$ and $\phi$, we can use the gradient in cylindrical coordinates:
\begin{eqnarray}
\Omega(R_0+ x, y)=&\Omega_0 +\left.\dfrac{\partial \Omega}{\partial R}\right|_{R=R_0} x & +\left.\dfrac{\partial \Omega}{R\partial \phi}\right|_{R=R_0} y  \\
=& \Omega_0 +\left.\dfrac{\partial \Omega}{\partial R}\right|_{R=R_0} x. &
\end{eqnarray}

We expand the velocity at first order in $x$ and $y$,
\begin{eqnarray}
v_x=& -(y_0+y)\Omega \approx & -y\Omega _0 \\
v_y=& (R_0+x)\Omega \approx &  R_0\Omega_0 +\left(\Omega_0 + R_0 \left.\dfrac{\partial \Omega}{\partial R}\right|_{R_0} \right)x
\end{eqnarray}
Using this, we can compute the local angular momentum in the z-axis, $L_z = \hat{z}\pmb{L}$, where $\pmb{L}$ is the angular momentum vector and $\hat{z}$ is the unit vector in the z-axis:
\begin{eqnarray}
L_z  = &\hat{z}\int \rho\ \pmb{r}\times\bvel\   d^3\pmb{x} \\
 = &\displaystyle \int \left[R\Omega_0 x +x^2\left(\Omega_0 +R\left.\dfrac{\partial \Omega}{\partial R}\right|_{R_0} \right) +y^2\Omega_0\right]\rho d^3\pmb{x} \\
 = & MR\Omega_0 x_{CM} +I_{11}\Omega_0\left(1+\left.\dfrac{\partial \ln \Omega}{\partial \ln R}\right|_{R_0} \right) + I_{22}\Omega_0 \ , \ I_{11}= \displaystyle \int  \rho x^2   d^3\pmb{x} \ , \ I_{22}=\displaystyle \int  \rho y^2   d^3\pmb{x} \\
\end{eqnarray}

where $x_{CM}$ is the position of the center of mass in the $x$-axis, and $I_{11}$ and $I_{22}$ are the components of the moment of inertia tensor. If the coordinates $x$ and $y$ are centered on the center of mass and $I_{11}=I_{22}=\frac{1}{2}I$ the angular momentum is reduced to
\begin{equation}
L_z=\dfrac{1}{2}I \ \Omega_0\left(1+\left.\dfrac{\partial \ln \Omega}{\partial \ln R}\right|_{R_0} \right) = \dfrac{1}{2}I\omega
\end{equation}

\section{Random Fields: Velocity, Vorticity and Normalized Circulation}
\label{sec:random_fields}
In this Section we show how to create GRFs for the noncircular velocity component $\vno$, using a function $\mV(k)$ in Fourier space. 
Let $\mathbb{W}(\pmb{r})$ be a Gaussian white-noise field with $\mu=0$, and $\sigma=1$. From $\mathbb{W}(\pmb{r})$ we create a Gaussian random velocity field $v(\pmb{r})$ by 
\begin{equation}
v (\pmb{r}) = \mathcal{F}^{-1}\left(\mV (\pmb{k}) \mathcal{F}( \mathbb{W} (\pmb{r}))\right) = \int e^{2\pi i \pmb{k}\cdot\pmb{r}} \mV (\pmb{k})\ d\pmb{k} \int  e^{-2\pi i \pmb{k} \cdot \pmb{r'}}\ \mathbb{W}(\pmb{r'}) d\pmb{r'} 
\end{equation}
where $ \mV (\pmb{k})$ is an even and positive function in Fourier space that generates the field  $v (\pmb{r})$, and $\mathcal{F}$ is the Fourier transform \citep{Lang_11}. The velocity dispersion of this field is given by:
\begin{equation}
\label{eq:sigma_0}
\sigma ^2 _v= \int_{\kmin}^{\kmax} \mV(\pmb{k})^2 d\pmb{k} = 2\pi \int_{\kmin}^{\kmax}\mV(k)^2 k dk.
\end{equation}
This velocity field generates a vorticity field $\bomega=\nabla \times \bvel$. The Fourier transform $\mathcal{F}$ of a field $f(\pmb{r})$ is given by $\hat{f}(\pmb{k})=\mathcal{F}[f(\pmb{r})] = \int e^{-2\pi i \pmb{k}\cdot \pmb{r}} f(\pmb{r}) d\pmb{r}$. Then, in Fourier space $ \hat{\omega} (\pmb{k}) = -2\pi i \pmb{k} \times \hat{v} (\pmb{k})$. More directly, $\bomega$ will be generated by the function $\mathcal{W}(k)=2\pi \mV(k)k$.
At the smallest scales the dispersion of both fields $\omega$ and $v$ will be related through
\begin{eqnarray}
\label{eq:sigma_vort_omega}
\frac{\sigma_{\omega}^2}{\sigma_{v}^2} =\frac{\displaystyle\int_{k_{\rm min}}^{k_{\rm max}} \mathcal{W}(k)^2 k dk }{\displaystyle\int_{k_{\rm min}}^{k_{\rm max}} \mV(k)^2 k dk } = 4\pi^2\ \frac{\displaystyle\int_{k_{\rm min}}^{k_{\rm max}} \mV(k)^2 k^3 dk }{\displaystyle\int_{k_{\rm min}}^{k_{\rm max}} \mV(k)^2 k dk }.
\end{eqnarray}
Now that we have a way to relate $\omega(x,y)$ with the function $\mV(k)$ that generates the random velocity field, let us compute the normalized circulation $\gamma(x,y,\ell)$,
\begin{eqnarray}
\label{eq:omega_model}
\omega(\pmb{r})=& \mathcal{F}^{-1}\left(\mathcal{W} (\pmb{k}) \mathcal{F}( \mathbb{W} (\pmb{r}))\right) =& \int e^{2\pi i \pmb{k} \cdot \pmb{r}}\ \mathcal{W} (\pmb{k})\ e^{-2\pi i \pmb{k} \cdot  \pmb{r'}}\ \mathbb{W}(\pmb{r'})\ d\pmb{k} d\pmb{r'}  \\
\gamma (\pmb{r},\ell)=& \displaystyle \dfrac{1}{\ell ^2}\int^{x+\frac{\ell}{2}}_{x-\frac{\ell}{2}} \int^{y+\frac{\ell}{2}}_{y-\frac{\ell}{2}} \omega(\pmb{r''}) d\pmb{r''}=& \frac{1}{\ell ^2}  \int^{x+\frac{\ell}{2}}_{x-\frac{\ell}{2}} \int^{y+\frac{\ell}{2}}_{y-\frac{\ell}{2}} d\pmb{r''}  \int e^{2\pi i \pmb{k} \cdot  \pmb{r''}} \mathcal{W} (\pmb{k}) e^{-2\pi i \pmb{k}\cdot \pmb{r'}} \mathbb{W} (\pmb{r'}) d\pmb{k} d\pmb{r'}.
\end{eqnarray}
Changing $\pmb{r''} \rightarrow \pmb{r}+\pmb{r''}$, we get
\begin{equation}
\gamma (\pmb{r},\ell)= \frac{1}{\ell ^2}  \int^{\frac{\ell}{2}}_{-\frac{\ell}{2}} \int^{\frac{\ell}{2}}_{-\frac{\ell}{2}} d\pmb{r''}  \int e^{2\pi i \pmb{k}\cdot \pmb{r}}  e^{2\pi i \pmb{k}\cdot \pmb{r''}} \mathcal{W} (\pmb{k}) e^{-2\pi i \pmb{k}\cdot \pmb{r'}} \mathbb{W} (\pmb{r'}) d\pmb{k} d\pmb{r'}
\end{equation}
We can integrate the term $ e^{2\pi i \pmb{k}\cdot \pmb{r''}}  d\pmb{r''} $ over the rectangular square
\begin{equation}
\int^{\frac{\ell}{2}}_{-\frac{\ell}{2}} \int^{\frac{\ell}{2}}_{-\frac{\ell}{2}}  e^{2\pi i \pmb{k}\cdot \pmb{r''}} d\pmb{r''} 
=\int ^{\frac{\ell}{2}}_{-\frac{\ell}{2}} e^{2\pi i k_x x''} dx'' \int ^{\frac{\ell}{2}}_{-\frac{\ell}{2}} e^{2\pi i k_y y''} dy''= \frac{\sin (\pi k_x \ell)}{\pi k_x} \frac{\sin (\pi k_y \ell)}{\pi k_y} 
\end{equation}
Now $\gamma(\pmb{r},\ell)$ has the form $\mathcal{F}^{-1}\left(M (\pmb{k}) \mathcal{F}( \mathbb{W} (\pmb{r}))\right)$

\begin{eqnarray}
M(\pmb{k})=\frac{1}{\ell ^2} \mathcal{W}(\pmb{k})\ \frac{\sin (\pi k_x \ell) \sin (\pi k_y \ell)}{\pi^2 k_x k_y}
\end{eqnarray}
Then, $\gamma(\pmb{r},\ell)$ is a random field in two dimensions, and its variance is given by

\begin{equation}
\label{eq:sigma_equation}
\sigma_{\gamma,\ell}^2 = \int M(\pmb{k})^2 d\pmb{k} = \frac{1}{\ell ^4}\int \mathcal{W}(\pmb{k})^2\  \frac{\sin (\pi k_x \ell)^2 \sin (\pi k_y \ell)^2}{ \pi^4 k_x^2 k_y^2}\ d k_x dk_y.
\end{equation}

This formulation has been tested by creating two-dimensional arrays from $\mV(k)$ representing the velocity field. From these simulated fields we calculate empirically $\sigma_{\gamma,\ell}^2$ which is in agreement with the equation derived in this Section. 

\section{Examples and Toy Models}
\label{app:examples}
In this Section we use toy models to show how the parameters of the $\mV(k)$ change the distribution of $\gamma$ at different scales $\ell$. For these models, we set $\sigma_{\omega}=1$ (Equation \ref{eq:sigma_vort_omega}), the box size $L$=1 and a resolution $L/N$ with $N=10^4$.  We use a function $\mV(k)$ of the form

\begin{equation}
\label{eq:power_spectrum}
\mV(k)\propto \left\{
\begin{array}{@{\,}r@{}ll}
 \quad &k^{-n_1}\quad &\text{ if } k_{\rm min}<k< k_{\rm c}\\ 
 \quad & k^{-n_2}\quad &\text{ if } k_c <k< k_{\rm max}\\
 \quad & 0 \quad \quad &\text{elsewhere} 
\end{array}\right.
\end{equation}

\begin{figure*}
\label{fig:circulation_random_fields}
\includegraphics[width=\linewidth]{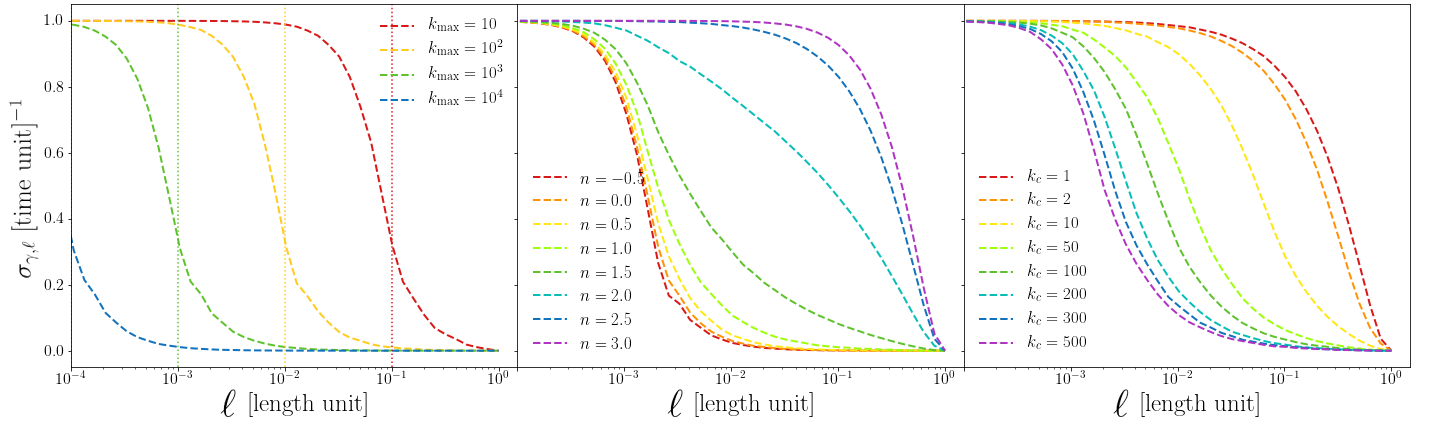}
\caption{Dispersion of the distribution of $\gamma(\pmb{r}, \ell)$ as a function of spatial scale. Left: The parameters $n_1$ and $n_2$ are fixed to zero, and $k_{\rm min}=1$. The parameter $\kmax$ is varied with values $\kmax \in [10,10^2,10^3,10^4]$. Middle: We fix the parameters $k_{\rm min}=1$ and $k_{\rm max}=500$. The parameters $n_1$ and $n_2$ are equal to $n_1=n_2=n$, which corresponds to a function $\mV(k)$ with a single power-law. The parameter $n$ takes values from the set $ [-0.5,0.0,0.5,1.0,1.5,2.0,2.5,3.0]$. Right: The fixed parameters are $k_{\rm min}=1$, $k_{\rm max}=500$, $n_1=1.0$, $n_2=2.5$. We vary the parameter $k_c \in [1,2,10,50,100,200,300,500]$}
\end{figure*}

We show the dispersion in $\gamma_{\ell}$ as a function of $\ell$ for different choices of $\mV(k)$ in Figure \ref{fig:circulation_random_fields}. The dispersion is calculated by the integral (\ref{eq:sigma_equation}). In the left panel we vary $\kmax$ while $n_1=n_2=0$, which represents white noise. The parameter $\kmax$ displaces the curve as a function of scale. In the middle panel we vary $n_1=n_2=n$ from -0.5 to 3.0 while $k_{\rm min}=1$ and $k_{\rm max}=500$. Fields with exponents between -0.5 and 1 show similar profiles. There is a notorious degeneracy between spectra with low exponents. In the right panel $k_{\rm min}=1$, $k_{\rm max}=500$, $n_1=1.0$, $n_2=2.5$, and we vary $k_c$ from $\kmin$ to $\kmax$.

\subsection{Toy Models}
\label{app:toy}
To illustrate the behavior of the normalized circulation distribution, we create toy models with different velocity fields. First, we define the circular velocity fields, which are given by
\begin{equation}
\bvel_{\rm rot}=R\Omega(R) \hat{\phi} \qquad  \qquad \Omega(R) \propto (R+r_0)^{-\beta}
\end{equation}

\begin{figure}
\label{fig:circulation_model_examples}
\includegraphics[width=\linewidth]{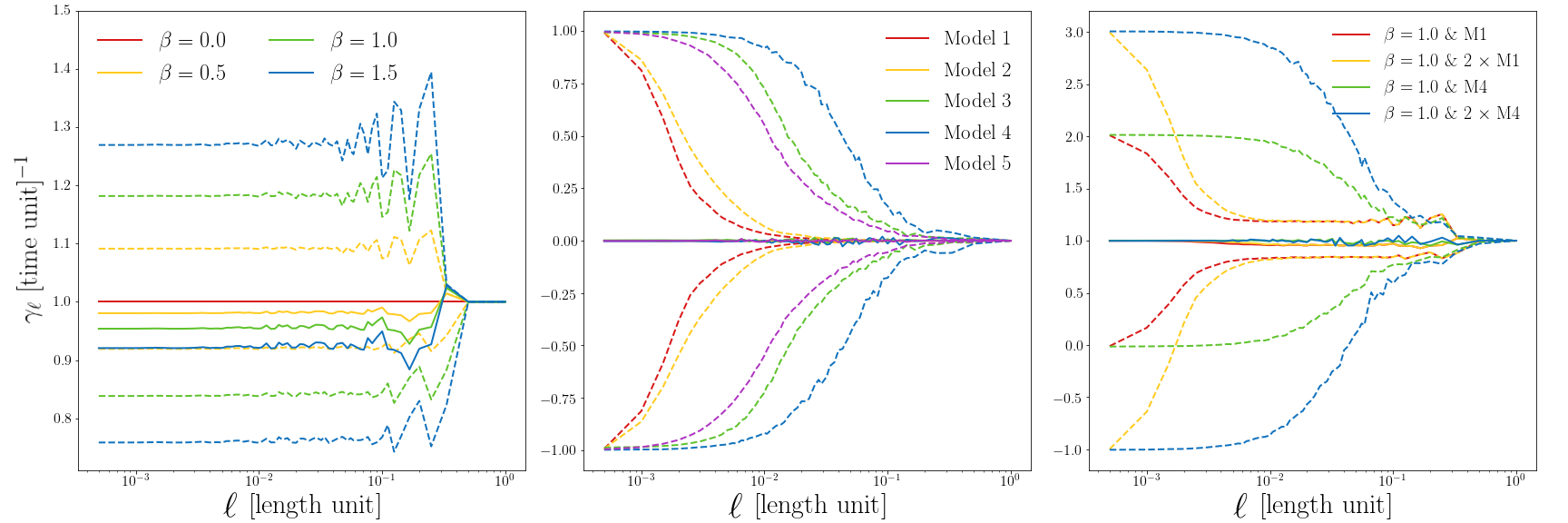}
\caption{ Percentiles of $\gamma$ for coherent galactic rotation, $\grot$, and for random velocity fields, $\gno$ and different toy models. Each field has been computed in a $2000 \times 2000$ grid. Left: solid lines show median values for $\grot$ as a function of the scale $\ell$, while dashed lines correspond to the 16th and 84th percentiles. Middle: percentiles of $\gno$ as a function of $\ell$, for different models of Gaussian random fields defined in Table \ref{table:random_parameters}. Dashed lines show the 16th and 84th percentiles, which correspond to 1$\sigma$ uncertainties for $\gno$. Solid lines show the median values of $\gno$. Right: Percentiles of toy models for $\grot+\gno$. The model of $\grot$ has $\beta=1$ for the four lines. The red and green lines correspond to $\grot$ plus model 1 and model 4, respectively. Yellow and blue lines have the same models for $\gno$ but with twice the magnitude. }
\end{figure}

The toy models are computed over a $2000\times 2000$ grid with a box size $L=1$, and we choose $r_0=1/2000$. The left panel of Figure \ref{fig:circulation_model_examples} shows the 16th, 50th, and 84th percentiles of $\grot$, the circulation of the circular velocity field, as a function of scale $\ell$. We normalize the values of $\grot$ such that the mean value of $\grot$ equals 1 at the highest resolution. We shows percentiles for $\beta \in [0.0,0.5,1.0,1.5]$. For solid rotation $\beta=0.0$ the distribution of $\grot$ is single valued. For other values of $\beta$ each percentile converges to an specific value at small $\ell$. Since $\Omega(R)$ is a decreasing function of radius, each percentile corresponds to $\grot$ measured along a unique radius in the field. It is important to notice that for these models $\grot$ is always positive. Variations at large-scale are due to low sampling.

In the middle panel we plot the percentiles for five models of Gaussian random velocity fields defined by their spectra $\mV(k)$ according to equation \ref{eq:power_spectrum}. The parameters of each model are shown in Table \ref{table:random_parameters}.

\begin{table}[ht]
\caption{Random field parameters}
\centering
\begin{tabular}{c c c c c c}
\hline\hline
Model & $n$ & $m$ & $k_{\rm min}$ & $k_{\rm max}$ & $k_c$\\ [0.5ex] 
\hline
1	&	0.0	&	-	&	4		&	500		&	- \\
2	&	1.0	&	-	&	64	    &	500		&	- \\
3	&	1.0	&	- 	&	4	    &	64		&	- \\
4 	& 	1.0 & 	3.0 & 	4		&	500		&	16 \\
5 	& 	1.0 & 	3.0 & 	4		&	500		&	64 \\ [1ex]
\hline
\end{tabular}
\label{table:random_parameters}
\end{table}
From these random velocity fields we obtain $\gno$ for each model. We normalize $\gno$ fields such that their variance $\sigma_{\gamma}^2=1$ at the highest resolution. Models 1 and 2 show similar distributions as a function of $\ell$. As seen before in Figure \ref{fig:circulation_random_fields}, at low values of the exponent $n$ the distribution of $\gno$ does not depend strongly on $n$. Model 3, a single power law from $k\in[4,64]$, lies close to model 5, which corresponds to a broken power law with $k_c=64$. The only difference between these two models is the behavior of $\mV(k)$ for values of $k > 64$: for model 3 $\mV(k)=0$ which is equivalent to $n_2=\infty$, while for model 5 $\mV(k)\propto k^{-3}$. This shows that functions $\mV(k)$ with high values of $n_2$ are similar to single power laws with $\kmax=k_c$.
The right panel of Figure \ref{fig:circulation_model_examples} shows the distribution of four different composite models using a velocity field with $\beta =1.0 $ plus models 1 and 4 times a factor of 1 or 2. Since model 1 changes mostly at small scales, for most of the spatial scales the circulation is given by $\grot$ until the width of both distributions is comparable. Then, if we increase the magnitude of the random field, the transition where $\Delta \grot \sim \Delta \gno$ moves to larger scales as shown by the yellow dashed lines in Figure \ref{fig:circulation_model_examples}. It is important to mention that as we increase the magnitude of the random field, the number of regions with negative circulation, i.e. with retrograde rotation with respect to the galaxy, also increases.

\section{Results with Different Models}

\subsection{Rotation curve}
\label{app:rotation}

Our method to compute $\orot$, and consequently $\grot$, is to measure a radial profile for the circular velocity field. The resulting radial profiles depend on the chosen size of the radial bins. In the main text we use the analytic function in equation \ref{eq:analytic} to compute $\grot$. To test the sensitivity with respect to the chosen rotation curves, we calculate $\lequi$ for different velocity models. The simplest model corresponds to measuring the median values of the rotation curve at intervals of 500 pc. To compute the derivatives in Equation \ref{eq:vorticity}, we fit a four-th order polynomial function to $\ln(v(R))$ as a function of $\ln(R)$. We show the resulting $\lequi$ as the blue line in Figure \ref{fig:test}.

We also test the effects of adding information in the azimuthal component by means of a Fourier series expansion and include the radial component of the velocity field. The velocity field and its vorticity are given by
\begin{eqnarray}
v_{\theta}(R,\theta,m)= A_0 + \sum_{j=1}^{m=4} A_j \cos(j\theta) + B_j \sin(j\theta) \qquad & \qquad
v_R(R,\theta,m)= C_0 + \displaystyle \sum_{j=1}^{m=4} C_j \cos(j\theta) + D_j \sin(j\theta)\\
\omega_z = \frac{1}{R} \left(\frac{\partial (R v_{\theta})}{\partial R} - \frac{\partial v_R}{\partial \theta}\right)
\end{eqnarray}
where the coefficients $A_j$, $B_j$, $C_j$, and $D_j$ are functions of radius and $m$ is the order of the expansion. To obtain $A_j(R)$, $B_j(R)$, $C_j(R)$, and $D_j(R)$ as functions of $R$, we separate the disk in radial bins of constant width. We tested bin widths of 200, 500 and 800 pc, but the led to almost the same results in $\lequi$. We show the effect of this complex velocity field on $\lequi$ in Figure \ref{fig:test} using a bin width of 200 pc. Since $\grot$ has more information about the circulation of the fluid, the effect of $\gno$ is being noticed at smaller scales.

\subsection{Different Models for the Random Velocity Field}
\label{app:power}

\begin{figure}
\centering
\label{fig:test}
\includegraphics[width=\linewidth]{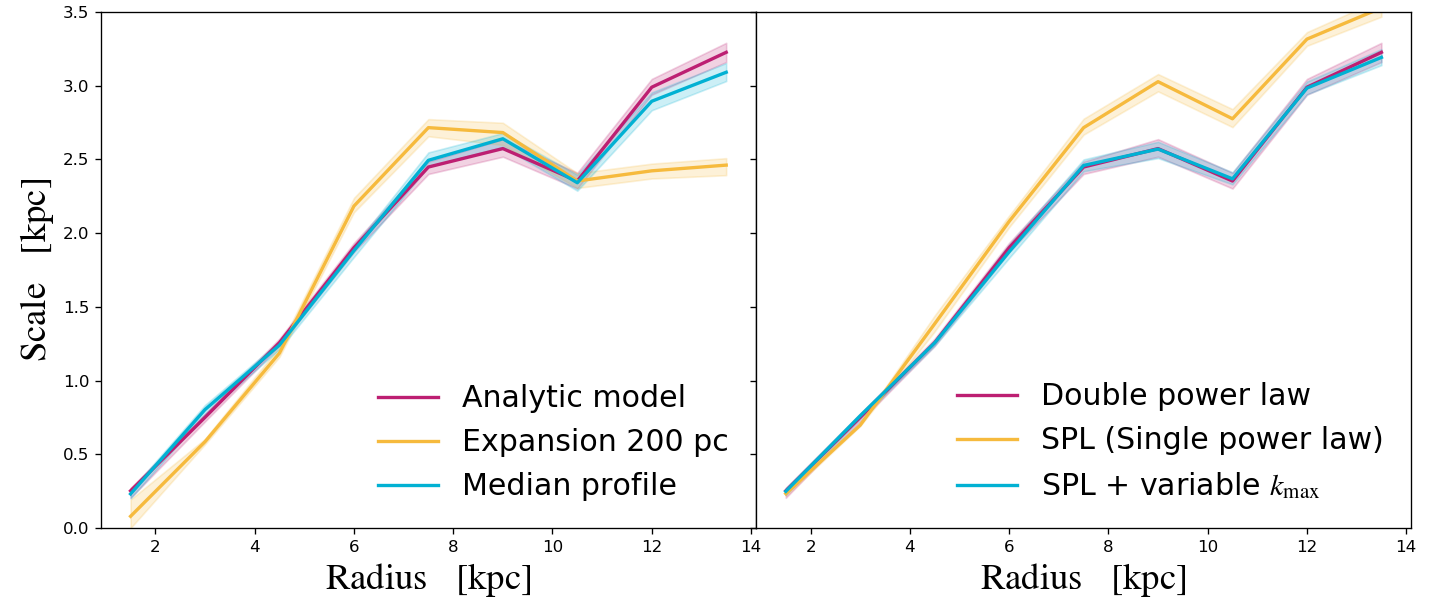}
\caption{Radial profiles of $\lequi$ {\it Left:} different choices of the large-scale velocity field. The magenta line shows $\lequi$ obtained in the main text, with its respective 1$\sigma$ uncertainties as a shaded region. The yellow line shows the resulting profile of $\lequi$ using a Fourier series expansion for the large-scale velocity field with a radial bin of 200 pc. The blue line shows the results for the median rotation curve.  {\it Right:} different models for $\mV(k)$. The magenta line corresponds to the model of $\mV(k)$ used in the main text. The model with a unique power law and fixed $\kmax$ is shown in yellow. The blue line corresponds to a single power law and variable $\kmax$.}
\end{figure}

\begin{figure}
\centering
\label{fig:distributions_example}
\includegraphics[width=0.75\linewidth]{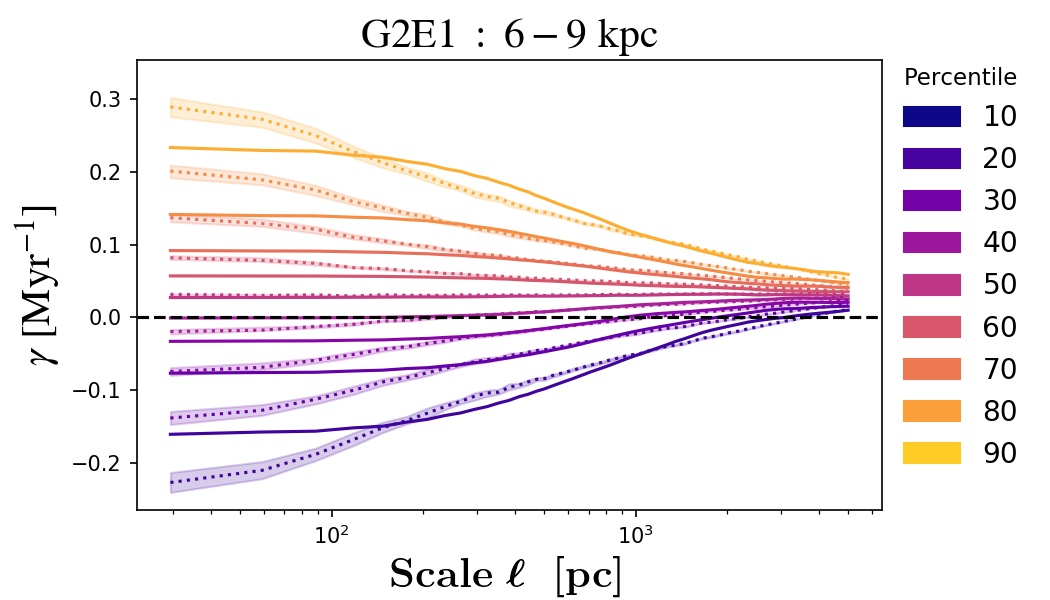}
\caption{Percentiles of $\gamma = \grot +\gno$ as a function of the scale $\ell$, for a single power law $\mV(k)$ from $\kmin=4/L$ to $\kmax=1/4\Delta x$. Solid lines show the percentiles of $\gamma$ measured in the simulations. Dotted lines show the percentiles of $\gno$ and the shaded regions correspond to 1$\sigma$ uncertainties.}
\end{figure}

For our model, the spatial scale $\lequi$, at which large-scale motions and noncircular motions contribute equally to the circulation of gas, depends on the function $\mV(k)$. We test different choices for $\mV(k)$. The first model of $\mV(k)$ consists of a power law ($n_1=n_2$), and the wavenumber $k$ is bounded between the values $\kmin=4/L$ and $\kmax=1/4\Delta x$. The second model also consists on a single power law, but allowing $\kmax$ to vary. We show these models in the right panel of Figure \ref{fig:test}. We can see that the model used for $\mV(k)$ in the main text shows similar results for $\lequi$ if we use a single power law with a variable $\kmax$. However, if we fix $\kmax=1/4\Delta x$, the values of $\lequi$ are higher. We show in Figure \ref{fig:distributions_example} an example of the percentiles obtained by using a function $\mV(k)$ with a single power law from $\kmin=4/L$ to $\kmax=1/4\Delta x$. We see that without the inclusion of the break at $k_c$ or a variable $\kmax$ the model is unable to reproduce the plateau of the percentiles at the smallest scales.

\section{Low Feedback Simulations}
\label{app:simulations}

In this Section we show the results for the simulations without early stellar feedback. Since these simulations have a lower stellar feedback, we add the suffix "-low" to distinguish them from the simulations presented in the main text.

Runs G2E1, G1E1, and G1E0.5 have 2.9, 1.3, and 1.4 times more gas than G2E1-low, G1E1-low, and G1E0.5-low, respectively. Particularly, G2E1 has $1.8\times10^{10} M_{\odot}$ of gas in the disk, while all the other runs have below $1\times10^{10} M_{\odot}$ in gas. This makes G2E1 the most unstable disk at this point of time. This might explain why G2E1 shows the largest values of $\sigma_0$, which translates into higher values of $\lequi$.

\begin{figure*}
\centering
\label{fig:dist_app}
\includegraphics[width=\linewidth]{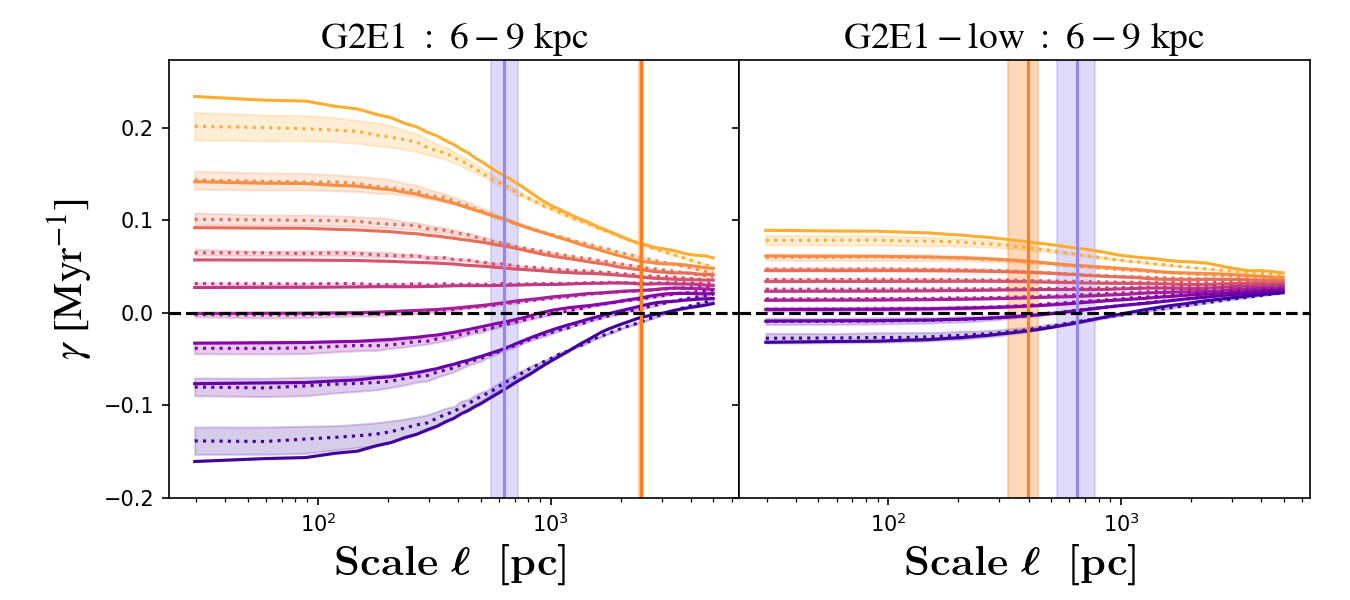}
\caption{Percentiles of circulation $\gamma$ within 6-9 kpc for G2E1 and G2E1-low, as a function of scale $\ell$. Solid lines represent the percentiles of $\gamma$ in the simulation, while dashed regions represent 1$\sigma$ uncertainty intervals for the model $\mV(k)$ around the median values, showed as dotted lines. Vertical lines show the spatial scale $\lk=1/\kc$ and its corresponding uncertainty illustrated by the dashed region. Black dashed horizontal lines show $\gamma=0$.}
\end{figure*}

If we compare the distribution of G2E1 and G2E1-low in Figure \ref{fig:dist_app}, the former shows a broader distribution of $\gamma$ and a slightly higher fraction of regions with retrograde rotation. However, as pointed out before, the comparison is not straightforward since G2E1-low has about a half the mass in gas compared to G2E1. 

\begin{figure*}
\centering
\label{fig:equi_app}
\includegraphics[width=\linewidth]{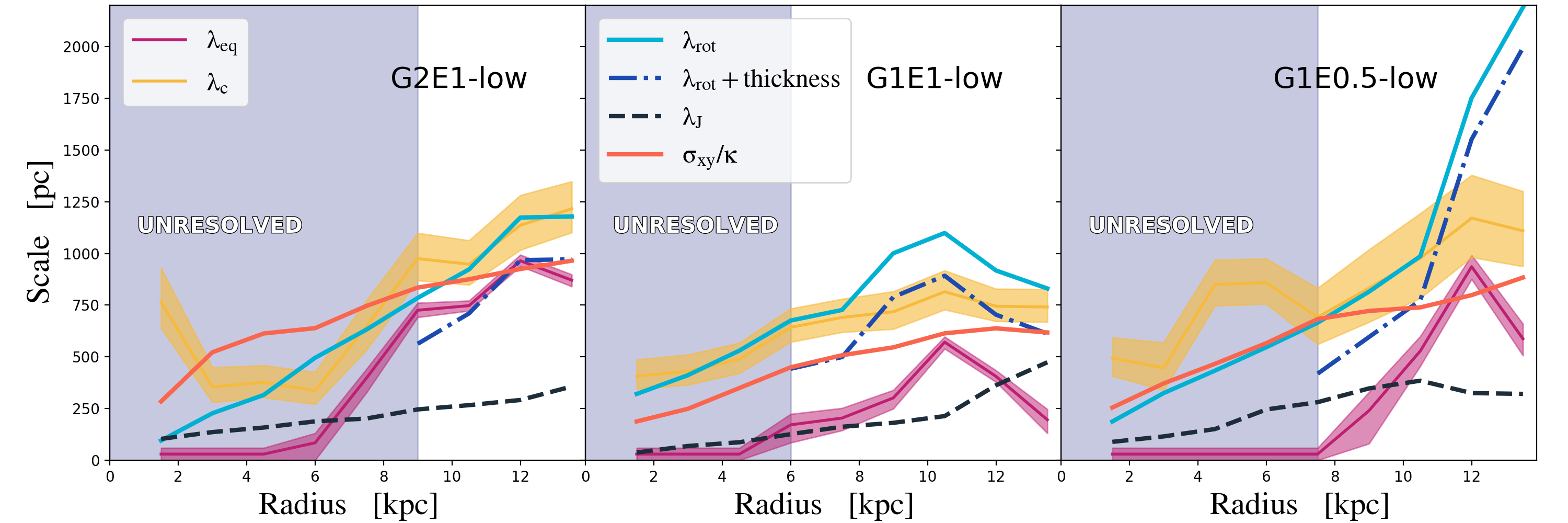}
\caption{Spatial scales as a function of galactocentric radius for simulations without early stellar feedback. Solid pink and yellow lines correspond to $\lequi$ and $\lk$, respectively. The shaded regions correspond to 1$\sigma$ uncertainties. The classical instability scales, $\lrot$ and $\lj$ are shown as a solid light-blue line and a black dashed line, respectively. The dotted dashed line shows the effects of the spatial resolution of the simulation on $\lrot$.}
\end{figure*}

In Figure \ref{fig:equi_app} we show the scales $\lequi$, $\lk$, and the scales of gravitational instability for G2E1-low, G1E1-low, and G1E0.5-low.

Each galaxy shows a similar behavior; at some particular radii $R^*$, $\lequi$ decreases beyond the resolution of the simulations. We see from Figures \ref{fig:rotation_curve} and \ref{fig:parameter_profile} that $\orot(R)$ decays exponentially while $\sigma_0$ decays somewhat linearly, which also holds for these simulations. At the center, where $\orot$ peaks, $\lequi \ll \Delta x$. Since $\orot$ decays faster, at some particular radius $\lequi \simeq \Delta x$, and $\lequi$ starts to be resolved.

The regions where $\lequi$ is resolved overlap with the regions where $\lrot$ is resolved, once we consider the effect of the disk thickness. This shows that $\lrot$ must be resolved in order to study $\lequi$ in simulations.

\newpage

\software{yt \citep{Turk_11},
Enzo \citep{Bryan_14}}

\bibliographystyle{yahapj}
\bibliography{bibliography}

\end{document}